\documentclass[floats,aps,prfluids,superscriptaddress,floatfix,onecolumn]{revtex4-2}

\usepackage{amssymb,amsmath}
\usepackage{csquotes}
\usepackage{siunitx}
\usepackage{graphicx}
\usepackage{hyperref}

\usepackage[usenames,dvipsnames]{color}
\usepackage[normalem]{ulem}

\renewcommand{\vec}[1]{\mathbf{\boldsymbol{#1}}}

\begin{document}

\title{Gradient dynamics model for drops of volatile liquid on a porous substrate}

\author{Simon Hartmann}
\email{s.hartmann@uni-muenster.de}
\thanks{ORCID ID: 0000-0002-3127-136X}
\affiliation{Institut f\"ur Theoretische Physik, Universit\"at M\"unster, Wilhelm-Klemm-Str.\ 9, D-48149 M\"unster, Germany}
\affiliation{Center for Nonlinear Science (CeNoS), Universit\"at M\"unster, Corrensstr.\ 2, 48149 M\"unster, Germany}

\author{Uwe Thiele}
\email{u.thiele@uni-muenster.de}
\homepage{http://www.uwethiele.de}
\thanks{ORCID ID: 0000-0001-7989-9271}
\affiliation{Institut f\"ur Theoretische Physik, Universit\"at M\"unster, Wilhelm-Klemm-Str.\ 9, D-48149 M\"unster, Germany}
\affiliation{Center for Nonlinear Science (CeNoS), Universit\"at M\"unster, Corrensstr.\ 2, 48149 M\"unster, Germany}

\begin{abstract}
  We present a mesoscopic hydrodynamic model for a spreading drop of volatile partially wetting liquid on a solid porous layer of small thickness. Thereby, evaporation takes place under strong confinement, i.e., we consider a drop that spreads on one of two parallel plates that form a narrow gap. Our gradient dynamics model describes the coupled dynamics of the vertically averaged vapor density profile, the drop height profile, and the vertically averaged saturation profile in the porous layer. The underlying free energy incorporates saturation-dependent wettability, capillarity of the drop and within the porous layer, air and vapor entropy, and an entropic contribution relevant close to complete filling of the porous layer. After developing the model we discuss the resulting sorption isotherm and illustrate typical drop spreading and imbibition behavior including the formation of a saturation halo within the porous substrate.
\end{abstract}

\maketitle

\section{Introduction}\label{sec:porous}

The behavior of liquids in and on porous media has been studied for a long time~\cite{Rich1931p,SaTa1958prslsa,BeJo1967jfm,Whit1986tpm,Marm1988jcis,DaHo1999pf} due to the wide spectrum of natural and technical systems where this is relevant. Examples include the motion of particles with a porous outer layer in a liquid~\cite{VaSF1996cj}, mushy regions occurring during solidification of alloys~\cite{Wors1992jfm}, droplets of simple or complex liquids spreading on and within porous solids~\cite{Marm1988jcis,DaHo1999pf,DaHo2000pf,SZKS2003acis,AlRa2004ces,XiSA2012l,Gamb2014cocis,AKTK2017l,YAHA2017jcp,JoTS2019ci,EDSD2021m}, reactive drops within a porous medium~\cite{Genn1999crasi}, the dewetting of a liquid film on a porous medium~\cite{ArRG2000epje,BaBr2001el}, thermocapillary instabilities in systems involving porous layers~\cite{Wors1992jfm,DeLH2001pre,SaUJ2010ces}, the infiltration of water into soil~\cite{CuJu2008prl,CuJu2009wrr,CuJu2009pre}, the coating of porous substrates~\cite{DeJZ2007jfm,GSMM2017jctr}, drops on polymer layers and brushes~\cite{EDSD2021m}, and phase behavior in nanoporous materials without~\cite{Hube2015jpcm} and with light-switched properties~\cite{WHJC2015pccp}.

Particular interest has focused on liquid flow past a porous medium~\cite{BaCh1995sjma,OcWh1995ijhmt,OcWh1995ijhmtb,Jeon2001pf,GLGV2003ijhmt,VaGO2007ces,ThGV2009pf,SaUJ2010ces,KWDM2015ijam,AnGO2017pre} as well as on the motion and stability of moving liquid-gas interfaces (imbibition fronts) within a porous medium~\cite{FiWo1994n,APBP1998pre,CuJu2008prl,CiIS2010pre,XiSA2012l,ChWY2015jpcc,BeCa2022el} including cases where chemical reactions occur~\cite{WiHo1999pf,DeW2001prl}. Also the absorption of vapors into porous materials is of interest~\cite{DoDT2003l}. Note finally that the creation of liquid-infused slippery substrates is often based on porous substrate layers~\cite{SaGa2014p,GWXM2015l,MMTM2016an,LGGM2017pra,BMWB2017apl}.

There are various approaches to modeling the dynamics of the imbibition process, i.e., the flow of liquid into and within porous material. In one approach the wet and dry regions within the porous medium are separated by a sharp imbibition front and the flow within the ``wet'' part is described via Darcy's law \(\vec{v}=-(\kappa/\eta)\nabla p\). It describes the velocity field \(\vec v\), which is driven by the pressure gradient \(\nabla p\) between the liquid reservoir and the Laplace pressure inside the pores of the porous medium~\cite{XiSA2012l, Gamb2014cocis}. Here, \(\kappa \) denotes the permeability of the porous medium, and \(\eta \) the dynamic viscosity of the liquid.

A spreading droplet on a permeable membrane (porous medium with only vertical pores) is modeled by~\citet{DaHo1999pf} using a thin-film equation with a sink term that is driven by the same pressure as the convective flux (see their section~II). In particular, their \mbox{Eqs.~(1a--c)} can be written as a gradient dynamics model~\cite{Thie2010jpcm} for the film height \(h\):
\begin{equation}
    \partial_t h \,=\,
    \nabla\cdot\left[Q\nabla\frac{\delta
            \mathcal{F}}{\delta h}\right]
    - M\left(\frac{\delta \mathcal{F}}{\delta h} - p_\mathrm{vap}\right),
    \label{eq:porous-onefield:gov}
\end{equation}
with the mobilities \(Q=h^3/3\eta \), \(M=\kappa/\eta \), the free energy \(\mathcal{F}=\int [\sigma (\partial_x h)^2/2+\varrho_\mathrm{liq} g h^2/2] \mathrm{d} x\), and \(p_\mathrm{vap}=0\). Here, \(\sigma \) is the interface tension, \(\varrho_\mathrm{liq}\) is the mass density of the liquid, and \(g\) is the gravitational acceleration. In this model, no lateral motion within the porous layer is accounted for.

\citet{DaHo1999pf} also describe in their section~III a model for a droplet on a fully saturated porous substrate where lateral Darcy flows may occur. Furthermore, a model with an explicit wet-dry boundary within the porous layer is considered (their section~IV). The process of imbibition into the vertical pores is included in Ref.~\cite{DaHo2000pf}, resulting in the coupled dynamics of height profiles on top and inside the porous medium, modeled through two coupled thin-film equations.
Both height profiles of the form \(h_i(x,\,t)\) are measured from the porous-liquid interface, i.e.,
every pore consists of a filled and an empty section.
This approach does not allow for lateral transport within the porous layer and is also followed by~\citet{AlRa2004ces}.

Partially filled pores can be accounted for in phase-field models of infiltration~\cite{CuJu2008prl, CuJu2009pre, CuJu2009wrr, CuJu2009jcp, CuJu2010wrr}. The governing equation describes the dynamics of a filling ratio within the porous medium and is mathematically closely related to a thin-film equation. Such models amend Richards equation~\cite{Rich1931p}, which itself extends Darcy's law to unsaturated porous media.
In a related approach,~\citet{KHHB2023jcp} have recently presented a gradient dynamics model for a volatile droplet on a grafted polymer substrate, which describes the coupled dynamics of the droplet, the liquid within the substrate and the vapor, also cf.~the review in Ref.~\cite{HDGT2024l}.

Here, we adapt the latter approach by replacing the polymer brush substrate with a porous medium of small finite thickness, and by furthermore incorporating the driving by capillary pressure as employed by~\citet{CuJu2009pre,BeCa2022el}. The resulting model effectively extends Eq.~\eqref{eq:porous-onefield:gov} with additional equations for the vapor dynamics and the dynamics of the liquid within the pores. It can be applied to studies of spreading drops of volatile liquids on porous substrates that are initially unsaturated.
The pursued gradient dynamics approach to mesoscopic hydrodynamics is based on the observation that in the Stokes limit, i.e., without inertia, the dynamics of a hydrodynamic system is simply driven by the minimization of a free energy functional. Such a thermodynamically consistent approach has for a long time been employed for the description of diffusive processes like the decomposition of binary mixtures via the Cahn-Hilliard equation~\cite{Cahn1961am,Cahn1965jcp,Bray1994ap}, before it was noted by~\citet{Mitl1993jcis} that the long-wave equation~\cite{OrDB1997rmp}, that describes the dewetting of a thin layer of simple nonvolatile liquid, belongs to the same class of models~\cite{Thie2010jpcm}. Subsequently, the finding was extensively employed in the analysis of such systems and, more importantly, extended to more complex thin-film systems~\cite{PBMT2005jcp,BCJP2013epje,ThTL2013prl,ThAP2016prf,HEHZ2022prsa,HDGT2024l}. Writing the hydrodynamic thin-film equations in gradient dynamics form automatically ensures that the derived equations respect the principles of thermodynamics, i.e., provides a convenient consistency check. This facilitates, in particular, the incorporation of additional effects via inclusion of further degrees of freedom and energy contributions~\cite{Thie2018csa}. 

The mesoscopic hydrodynamic model is presented in section~\ref{sec:model} as a three-field gradient dynamics. Particular attention rests on the involved transport processes and the underlying energy functional. The subsequent section~\ref{sec:sorption} discusses the resulting sorption isotherm of the porous medium, while section~\ref{sec:spreading} analyzes the coupled spreading, evaporation and imbibition dynamics of a sessile drop on a porous substrate. Finally, we conclude in section~\ref{sec:conc} with a brief discussion.

\section{Mesoscopic hydrodynamic model}
\label{sec:model}

\subsection{Three-Field Gradient Dynamics}

\begin{figure}[!htb]
    \centering
    \includegraphics[width=0.8\hsize]{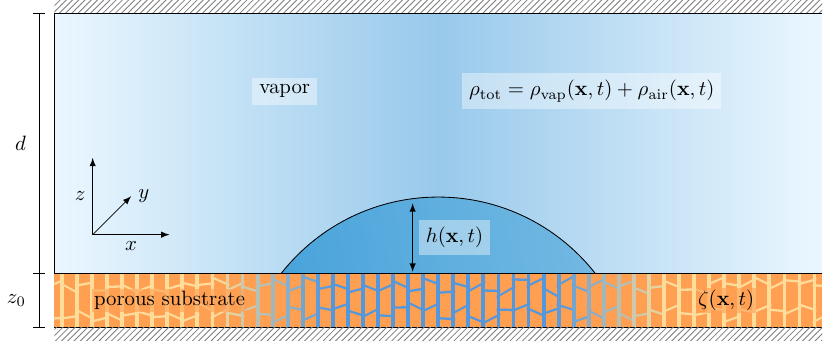}\caption{Sketch of the considered geometry for a volatile liquid drop on a porous substrate within a shallow chamber of height \(d\). The porous substrate has mostly vertical pores with some horizontal connections allowing for (slow) lateral diffusion of liquid. The drop profile is described by the height \(h(\mathbf x, t)\), the substrate has a constant thickness \(z_0\), and the liquid contained in the pores is described by an effective height \(\zeta(\mathbf x, t)\). The particle densities of vapor \(\rho_\mathrm{vap}(\mathbf x, t)\) and ambient air \(\rho_\mathrm{air}(\mathbf x, t)\) together account for a constant total density \(\rho_\mathrm{tot}\) in the gas phase.}\label{fig:porous-sketch}
  \end{figure}
  
The gradient dynamics model that we develop in the following includes fields for the local vapor concentration above the substrate and the local amount of liquid within the porous medium. A sketch of the considered geometry is shown in Fig.~\ref{fig:porous-sketch}.
Importantly, we assume that both, the gas layer and the porous substrate, are thin as compared to lateral extensions such that the local vapor concentration, as well as the local filling ratio of the pores, are approximately homogeneous in vertical direction. For the vapor layer this is ensured by considering a confined geometry, i.e., the gas phase is limited by a horizontal solid plate situated at a distance \(d\) from the substrate. Such a setup for the description of the diffuse vapor dynamics was introduced for a volatile drop on a smooth solid substrate by~\citet{HDJT2023jfm} and employed for such a drop on a brush-covered substrate in Ref.~\cite{KHHB2023jcp}.

Given these assumptions, we use three variables \(\psi_i(\mathbf x, t)\) to describe the state of the system formed by the drop, porous substrate, and vapor layer.
The general gradient dynamics framework reads~\cite{Thie2018csa}
\begin{equation}
    \partial_t \psi_i = \nabla\cdot  \left[ \sum_{b=1}^3 \,Q_{ij}\,
    \nabla \frac{\delta \mathcal{F}}{\delta \psi_j}\right]
    - \sum_{j=1}^3 \,M_{ij}\, \frac{\delta \mathcal{F}}{\delta \psi_j}
    \label{eq:porous-three-field-gradient-dynamics}
\end{equation}
where the subscripts \(i,\,j = 1,\,2,\,3\) refer to the three fields and the \(3\times 3\) mobility matrices  with components $M_{ij}$ and $Q_{ij}$ represent the conserved and nonconserved dynamics, respectively.

The variable \(\psi_1\) describes the number of liquid particles within the drop per substrate area and relates to its profile height \(h\) as
\begin{equation}
    \psi_1(\vec x,\,t)= \rho_\mathrm{liq} h(\vec x,\,t).\label{eq:porous-psi1}
\end{equation}
where $\rho_\mathrm{liq} $ is the constant and uniform particle number density of the liquid.
Further, \(\psi_3\) describes the number of vapor particles within the gas layer per substrate area  and relates to the height-averaged vapor density \(\rho_\mathrm{vap}(\mathbf x, t)\) via
\begin{equation}
    \psi_3(\vec x,\,t)= \rho_\mathrm{vap}(\vec x,\,t) [d-h(\vec x,\,t)].
    \label{eq:porous-psi3}
  \end{equation}
For simplicity, we consider the system as isothermal and assume that the total particle density \(\rho_\mathrm{tot}\) in the gas layer is constant and uniform. This assumption reflects that the velocity of sound is much larger than the velocity of all dynamics processes we are interested in.

Here, \(\rho_\mathrm{tot}\) is the sum of the vapor particles and of the remaining particles of the atmosphere constituents, simply labeled as ``air''. Therefore,
\begin{equation}
    \rho_\mathrm{tot} = \rho_\mathrm{vap}(\vec x,\,t) + \rho_\mathrm{air}(\vec x,\,t).
\end{equation}

Finally, the variable \(\psi_2(\vec x,\,t)\) is the number of liquid particles within the porous layer per substrate area. It is proportional to the local pore filling ratio \(\Theta(\vec x,\,t)\), i.e.,
\begin{equation}
    \psi_2(\vec x,\,t)= \rho_\mathrm{liq} z_0 b \,\Theta(\vec x,\,t) = \rho_\mathrm{liq} \zeta(\vec x,\,t)
    \label{eq:porous-psi2}
\end{equation}
where \(b\) is the uniform and constant porosity (void fraction) and $z_0$ is the thickness of the porous layer. We have also introduced \(\zeta(\vec x,\,t)=z_0 b \,\Theta(\vec x,\,t)\) as the effective height of the substrate-contained liquid.

\subsection{Transport Processes}
The conserved mobilities \(\mathbf{\underline{Q}}\) account for three processes: (i) viscous motion within the drop, (ii) Darcy-type diffusive transport of liquid particles within the substrate, and (iii) diffusive transport of vapor particles within the gas phase. This automatically neglects any dynamic coupling between the phases.
In particular, there is no direct dynamic coupling between the liquid in the drop and the liquid in the porous layer, i.e., \(Q_{12}=Q_{21}=0\). This should be a good approximation at small porosity, but could in principle be improved along the lines of~\citet{ThGV2009pf}, i.e., by considering Darcy-Brinkman dynamics in the porous layer.
The resulting  mobility matrix is diagonal, namely,
\begin{equation}
    \mathbf{Q}
    =
    \begin{pmatrix}
        \frac{1}{\rho_\mathrm{liq}}\,\frac{\psi_1^3}{3 \eta} & 0                          & 0 \\
        0                                                    & \hat D_\mathrm{por} \psi_2 & 0 \\
        0                                                    & 0                          &
        \frac{1}{k_B T} D_\mathrm{vap} \psi_3
    \end{pmatrix}
    \label{eq:porous-mob-c}
\end{equation}
with the vapor diffusion coefficient \(D_\mathrm{vap}\). The mobility of the liquid within the porous substrate, \(\hat D_\mathrm{por}\), needs to be determined, e.g., from Darcy's equation~\cite{Whit1986tpm} and will depend on the permeability \(\kappa \) of and the effective viscosity \(\eta_\mathrm{eff}=\eta/b\) in the porous layer.
Following the phase-field approach of~\citet{CuJu2009pre}, we obtain
\begin{equation}
    \hat D_\mathrm{por} = z_0\kappa k_r(\Theta)/\eta_\mathrm{eff},
\end{equation}
where \(k_r(\Theta)\) is a relative permeability, i.e., an increasing convex function of the liquid saturation \(\Theta \). For an example of such a function see Fig.~2 of Ref.~\cite{CuJu2009pre}. Darcy's law is recovered for \(k_r(\Theta) = \text{const}\).

The nonconserved mobilities \(\mathbf{\underline{M}}\) account for all three transfer processes, namely, (i) evaporation (or condensation) between the film and the gas phase, (ii) evaporation (or condensation) directly between the substrate and the gas phase, and (iii) liquid transfer between the porous substrate and the drop. We assume that these processes are solely driven by the partial pressures of the phases and assign corresponding Onsager coefficients. The resulting matrix reads
\begin{equation}
    \mathbf{M} = \begin{pmatrix}
        M_\mathrm{im} + M_\mathrm{ev} & -M_\mathrm{im}                 & -M_\mathrm{ev}               \\
        -M_\mathrm{im}                & M_\mathrm{im} + M_\mathrm{ev}' & -M_\mathrm{ev}'              \\
        -M_\mathrm{ev}                & -M_\mathrm{ev}'                & M_\mathrm{ev}+M_\mathrm{ev}'
    \end{pmatrix}.
    \label{eq:porous-mob-nc}
\end{equation}
where \(M_\mathrm{ev}\) and \(M_\mathrm{ev}'\) are the Onsager coefficient of phase change for the drop and the substrate, respectively, and \(M_\mathrm{im}\) is an imbibition rate coefficient. The particular form of the matrix ensures that \(\psi_1+\psi_2+\psi_3\) is conserved, i.e., \(\partial_t (\psi_1+\psi_2 +\psi_3) = -\nabla\cdot\vec{j}_{\psi_1+\psi_2 +\psi_3}\).

As the porous medium exposes only a fraction of the contained liquid to the surface via the pores, we ensure that the evaporation rate from the substrate surface scales with the porosity employing the simple assumption
\begin{equation}
    M_\mathrm{ev}' = b M_\mathrm{ev}.
\end{equation}
Keeping the overall structure of the matrix, in principle, more complicated dependencies may be incorporated into the $M$'s.

\subsection{Free Energy Functional}

Next, we discuss the underlying free energy functional \(\mathcal{F}[\psi_1,\,\psi_2,\,\psi_3]\). We employ the long-wave approximation~\cite{OrDB1997rmp,Thie2007chapter} and, for convenience, use the fields \(h\), \(\rho_\mathrm{vap} \), and \(\zeta \) as abbreviations for the relations containing only $\psi_i$. We have
\begin{align}
    \mathcal{F} =&\int_\Omega \Bigg[ \underbrace{\frac{\gamma_\mathrm{lg}}{2} (\nabla h)^2}_\text{liquid-gas\,interface\,energy} + \underbrace{\gamma_\mathrm{sl}(\Theta)}_\text{liquid-substrate\,interface\,energy} + \underbrace{f_\text{wet}(h,\,\Theta)}_\text{wetting energy}+\underbrace{(h+\zeta) f_\text{liq}(\rho_\mathrm{liq})}_{\text{liquid bulk energy}}
        \nonumber\\
        &         +\,\underbrace{(d-h) f_\text{vap}(\rho_\mathrm{vap})}_{\text{vapor energy}}
        \,+\,\underbrace{(d-h) f_\text{air}(\rho_\mathrm{air})}_{\text{air energy}}
        \,+\,\underbrace{z_0\frac{\Gamma}{2}(\nabla\Theta)^2 \,+\, z_0\Psi(\Theta)}_{\text{substrate energy}} \Bigg] \mathrm{d} x \mathrm{d}y, \label{eq:porous-energy}
    \nonumber
\end{align}
where \(f_\mathrm{liq}\), \(f_\mathrm{vap}\), and \(f_\mathrm{air}\) are bulk liquid, vapor, and air energies per volume, \(\gamma_\mathrm{lg}\) is the constant liquid-gas interface energy, \(\gamma_\mathrm{sl}\) is the (adaptive) filling ratio-dependent liquid-substrate interface energy, and \(f_\text{wet}(h,\,\Theta)\) is the (per area) wetting energy.

The remaining terms (last two) represent the energetic description of the porous substrate layer that we base on the work of~\citet{CuJu2009pre}.
Namely, the substrate energy contains the \emph{capillary potential} \(\Psi(\Theta)\) and a nonlocal term that encodes the apparent interface energy \(\Gamma\) of imbibition fronts. The factor \(z_0\) transforms the per-volume contributions into a per-area contribution in our height-averaged setting. It also relates the filling ratio \(\Theta\) to the effective height of liquid in the pores, \(\zeta=z_0 b \Theta\), see Eq.~\eqref{eq:porous-psi2}.
The capillary potential is related to the capillary pressure function \(J(\Theta)\sim -\partial_\Theta \Psi\). Analytical expressions for \(J\) and the relative permeability \(k_r(\Theta)\) are given, e.g., by the van Genuchten-Mualem model (vGM)~\cite{Genu1980sssaj, CuJu2009pre}, where
\begin{align}
    k_r &= \sqrt{\Theta} \left[1 - \left(1-\Theta^{1/m}\right)^m\right]^2,\\
    J &= p_c \, \left(\Theta^{-1/m} - 1\right)^{1/n}.\label{eq:porous-vGM}
\end{align}
Here, $p_c$ is the scale of the capillary pressure, \(n>1\) is a measure for pore size distribution, and \(m=1-1/n\).
The capillary pressure constant $p_c$ results from the Laplace pressure inside the pores and relates to the capillary rise $h_\mathrm{cap}$ via Jurin's law~\cite{Juri1718pt, CuJu2009pre}, i.e.,
\begin{equation}
    p_c = \frac{2 \gamma_\mathrm{lg} \cos \theta_\mathrm{eq}}{r_0} = \varrho_\mathrm{liq}\,g\,h_\mathrm{cap}
\end{equation}
with the typical pore radius \(r_0 \approx \sqrt{\kappa/b}\), the liquid mass density $\varrho_\mathrm{liq}$, and the equilibrium contact angle \(\theta_\mathrm{eq}\) of the liquid on the pure (non-porous) substrate material.

\begin{figure}[!tb]
    \centering
    \includegraphics[width=0.8\hsize]{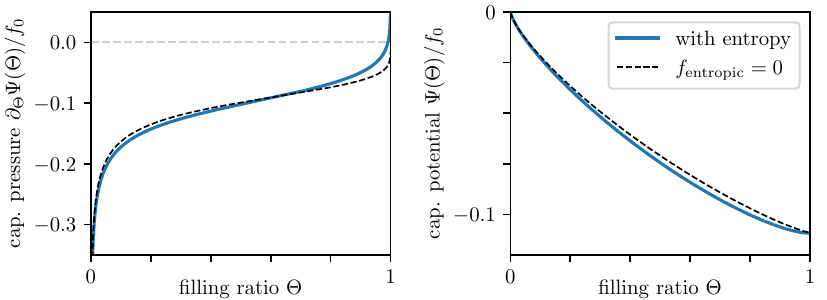}
    \caption{Capillary pressure \(\partial_\Theta \Psi\) and corresponding potential \(\Psi\) as functions of the pore filling ratio \(\Theta\) according to Eq.~\eqref{eq:porous-capillary-pressure} in units of \(f_0=\rho_\mathrm{liq} k_BT\). We consider a low porosity \(b=\num{0.01}\), all other parameters are as in Tab.~\ref{tab:porous-params}, i.e., the dimensionless capillary pressure constant is \(p_c / f_0 \approx \num{0.1}\). Shown are the cases with (solid line) and without (dashed line) the entropic correction \(f_\text{entropic}\), which ensures that the capillary pressure diverges when the pores are completely filled, \(\Theta\to 1\). Note that pressure and potential have identical units as $\Theta$ is dimensionless.}\label{fig:porous-energy}
  \end{figure}

Note that the capillary pressure \(\partial_\Theta \Psi\) is always negative for \(0<\Theta<1\), zero for \(\Theta=1\), and undefined for \(\Theta>1\).
In other words, the energetic minimum of the capillary potential is at full saturation \(\Theta = 1\). The case \(\Theta>1\) corresponds to the (unphysical) oversaturation of the porous medium with liquid, i.e., the volume of the contained liquid would be larger than the space available in the pores. To prevent such unphysical behavior in our dynamical model, we additionally incorporate and entropic contribution arising for the two-phase system formed by the liquid and the gas within the pores~\cite{Doi2013}. For the nearly saturated liquid of interest, this gives the entropic contribution to the local free energy
\begin{equation}
    f_\text{entropic}(\Theta) = b \, \rho_\mathrm{liq} \, k_BT \, (1-\Theta)\log(1-\Theta).\label{eq:porous-entropy}
\end{equation}
The complementary entropic contribution of the gas in the pores is negligible against the capillary potential that is part of the employed  vGM model~\eqref{eq:porous-vGM}.
We combine the entropic contribution~\eqref{eq:porous-entropy} with the previously employed vGM expression~\eqref{eq:porous-vGM} into the capillary potential \(\Psi\). Here, we use this to directly define the pressure
\begin{equation}
    \partial_\Theta \Psi = - J + \partial_\Theta f_\text{entropic}.\label{eq:porous-capillary-pressure}
\end{equation}
At low porosity \(b\), the entropic part only corresponds to a small correction to the capillary pressure (and resulting potential), as can be assessed in Fig.~\ref{fig:porous-energy}. There both the capillary pressure \(\partial_\Theta \Psi\) and potential \(\Psi\) are illustrated.
Nevertheless, including the entropy is crucial as it ensures that the capillary pressure diverges when approaching complete filling, \(\Theta\to 1\), i.e., the minimum of the capillary potential is always at \(\Theta < 1\).

To account for the adaptive wettability of the porous substrate we follow an approach similar to~\citet{KHHB2023jcp}. Namely, we assume a power law dependence of the Hamaker constant on the volume fraction \(c\) of the substrate material in the porous layer.
\begin{equation}
    f_\text{wet}(h,\,\Theta) = A_0\,c^a \left( \frac{h_p^3}{5h^5} - \frac{1}{2h^2} \right),
\end{equation}
where \(A_0\)  is the Hamaker constant of a poreless substrate, and \(a \geq 0\) is some constant exponent. The substrate volume fraction is \(c=1-b\Theta\), i.e., the complement to the liquid concentration.

Furthermore, we assume that the interface energy between the drop and the porous layer adapts with the same power law as used in the wetting potential. This ensures consistency with the adaption of a macroscopic substrate-gas interface energy, similar to the case of a polymer-brush substrate discussed in Ref.~\cite{GrHT2023sm}. Hence, we write for the substrate-liquid interfacial energy
\begin{equation}
    \gamma_\mathrm{sl}(\Theta) = \gamma_\mathrm{sl,\,dry}\,c^a
\end{equation}
with some positive exponent \(a\). This implies that the interface energy decreases when the substrate fills with liquid.

Vapor and air are considered to correspond to ideal gases, cf.~Ref.~\cite{HDJT2023jfm}. In consequence, their respective free energy densities are purely entropic
\begin{equation}
    f_\text{vap} = k_B T \rho_\mathrm{vap} \left[ \log(\Lambda^3\rho_\mathrm{vap}) - 1 \right]
    ~~\text{and}~~
    f_\text{air} = k_B T \rho_\mathrm{air} \left[ \log(\Lambda^3\rho_\mathrm{air}) - 1 \right]
\end{equation}
with the mean free path length \(\Lambda\). This completes the discussion of all modeling elements.

\subsection{The kinetic model}
The dynamic equations~\eqref{eq:porous-three-field-gradient-dynamics} involve the variations of the free energy functional, which compute as
\begin{equation}
    \begin{aligned}
        \frac{\delta \mathcal{F}}{\delta h} = \rho_\mathrm{liq} \frac{\delta \mathcal{F}}{\delta \psi_1} &= -\gamma_\mathrm{lg} \Delta h + \partial_h f_\text{wet}(h,\,\Theta) + f_\text{liq}\\
        \frac{\delta \mathcal{F}}{\delta \zeta} = \rho_\mathrm{liq} \frac{\delta \mathcal{F}}{\delta \psi_2} &= \partial_\zeta \left[ \gamma_\mathrm{sl}(\Theta) + f_\text{wet}(h,\,\Theta) + z_0 \Psi (\Theta) \right] + f_\text{liq} - \frac{\Gamma}{z_0 b} \Delta \Theta\\
        \frac{\delta \mathcal{F}}{\delta \psi_3} &= k_B T \log \left( \frac{\rho_\mathrm{vap}}{\rho_\mathrm{tot}-\rho_\mathrm{vap}} \right),
    \end{aligned}\label{eq:porous-pressures}
\end{equation}
where in the final equation we have used that the vapor particle density is much smaller than the total gas particle density, which itself is much smaller than the liquid density \({\rho_\mathrm{vap} \ll \rho_\mathrm{tot} \ll \rho_\mathrm{liq}}\), as discussed in more detail by~\citet{HDJT2023jfm}.

Inserting the energy variations as well as the mobility matrices (Eqs.~\eqref{eq:porous-mob-c}~and~\eqref{eq:porous-mob-nc}) into the three-field gradient dynamics Eq.~\eqref{eq:porous-three-field-gradient-dynamics} gives the resulting dynamic equations 
\begin{equation}
    \begin{aligned}
        \partial_t \psi_1 / \rho_\mathrm{liq} &= \nabla\cdot \left[ \frac{\psi_3^3}{3\eta \, \rho_\mathrm{liq}^2} \nabla \frac{\delta \mathcal{F}}{\delta \psi_1} \right] - j_\mathrm{ev} - j_\mathrm{im},\\
        \partial_t \psi_2  / \rho_\mathrm{liq} &= \nabla\cdot \left[ \frac{D_\mathrm{por}}{\rho_\mathrm{liq}} \, \psi_2 \, \nabla \frac{\delta \mathcal{F}}{\delta \psi_2} \right] - j_\mathrm{ev}' + j_\mathrm{im},\\
        \partial_t \psi_3 / \rho_\mathrm{liq} &= \nabla\cdot \left[ D_\mathrm{vap} (d-h) \nabla \rho_\mathrm{vap} \right] / \rho_\mathrm{liq} + j_\mathrm{ev} + j_\mathrm{ev}'.
    \end{aligned}
\end{equation}
The transfer fluxes are
\begin{align}
    j_\mathrm{ev} &= \frac{M_\mathrm{ev}}{\rho_\mathrm{liq}} \left(\frac{\delta \mathcal{F}}{\delta \psi_1} - \frac{\delta \mathcal{F}}{\delta \psi_3}\right),\\
    j_\mathrm{ev}' &= \frac{M_\mathrm{ev}'}{\rho_\mathrm{liq}} \left(\frac{\delta \mathcal{F}}{\delta \psi_2} - \frac{\delta \mathcal{F}}{\delta \psi_3}\right),\\
    j_\mathrm{im} &= \frac{M_\mathrm{im}}{\rho_\mathrm{liq}} \left(\frac{\delta \mathcal{F}}{\delta \psi_1} - \frac{\delta \mathcal{F}}{\delta \psi_2}\right),
\end{align}
namely, the film evaporation/condensation flux, the evaporation/condensation flux of the liquid contained in the porous layer, and the imbibition flux, respectively.

Finally, for an efficient numerical treatment, we reformulate the governing equation in terms of the dimensionless relative vapor concentration (or relative humidity) \(\phi(\vec x, t) = k_BT \rho_\mathrm{vap}(\vec x, t)/p_\mathrm{sat}\) and the dimensionless filling ratio of the substrate \(\Theta(\vec x, t) = \zeta / (z_0 b)\).
\begin{equation}
    \begin{aligned}
        \partial_t h &= \nabla \left[ \frac{h^3}{3\eta} \nabla \frac{\delta \mathcal{F}}{\delta h} \right] - j_\mathrm{ev} - j_\mathrm{im}\\
        \partial_t \Theta &= \nabla \left[ \frac{\kappa k_r(\Theta)}{\eta} \nabla \frac{\delta \mathcal{F}}{\delta \zeta} \right] + \frac{1}{z_0 b} (j_\mathrm{im} - j_\mathrm{ev}')\\
        \partial_t [(d-h)\phi] &= \nabla \left[ D_\mathrm{vap} (d-h) \nabla \phi \right] + \frac{\rho_\mathrm{liq} k_B T}{p_\mathrm{sat}} (j_\mathrm{ev} + j_\mathrm{ev}').
    \end{aligned}\label{eq:porous-model}
\end{equation}
This completes the model development. The model is now employed to discuss in the next section the resulting sorption isotherm of the porous medium. The subsequent section~\ref{sec:spreading} analyzes the coupled spreading, evaporation and imbibition dynamics of a sessile drop on a porous substrate. 
\section{Sorption Isotherm of a Porous Medium}
\label{sec:sorption}
Before we employ numerical tools to study the behavior of the coupled three-phase model, we consider spatially homogeneous equilibrium states, where \(h(\vec x, t)=\text{const}\), \(\Theta(\vec x, t)=\text{const}\), and \(\phi(\vec x, t)=\text{const}\).
These states can be assessed analytically, as the governing equations strongly simplify when all spatial and temporal derivatives are set to zero. In addition, we assume that the liquid film is either very thick or that it corresponds to a thin precursor layer, i.e., $h\gg h_p$ or  $h=h_p$, respectively. In both cases the wetting potential is negligible, \(f_\mathrm{wet}(h,\,\Theta)\approx 0\).

Evaluating the partial pressures (Eq.~\eqref{eq:porous-pressures}) for a homogeneous equilibrium state yields
\begin{equation}
    \begin{aligned}
        \frac{\delta \mathcal{F}}{\delta \psi_1} &= \frac{1}{\rho_\mathrm{liq}}f_\text{liq}\\
        \frac{\delta \mathcal{F}}{\delta \psi_2} &= \frac{1}{\rho_\mathrm{liq}} \left\{ \partial_\zeta \left[ \gamma_\mathrm{sl}(\Theta) + f_\text{wet}(h, \Theta) + z_0 \Psi(\Theta) \right] + f_\text{liq} \right\}\\
        \frac{\delta \mathcal{F}}{\delta \psi_3} &= k_B T \log \left( \frac{\rho_\mathrm{vap}}{\rho_\mathrm{tot}-\rho_\mathrm{vap}} \right).
    \end{aligned}\label{eq:porous-pressures_flat}
\end{equation}
In thermodynamic equilibrium, the system is at rest. This implies that the transfer fluxes \(j_\mathrm{ev}\), \(j_\mathrm{ev}'\), and \(j_\mathrm{im}\) are zero as the partial pressures balance. In contrast, a larger pressure in the film than in the substrate causes an absorption of liquid into the substrate until the pressures equilibrate.
Their balance results in coexistence conditions for the three phases.

When a thick homogeneous film is in equilibrium with an ambient vapor phase, the air is considered saturated with vapor.
Hence, comparing the film and vapor partial pressures defines the saturation vapor density, i.e.,
\begin{alignat}{2}
    &  & \frac{\delta \mathcal{F}}{\delta \psi_1} & = \frac{\delta \mathcal{F}}{\delta \psi_3}\Big|_{\rho_\mathrm{vap}=\rho_\mathrm{sat}}\\
    \Leftrightarrow\mkern40mu &  & \frac{f_\mathrm{liq}}{\rho_\mathrm{liq}} & = k_B T \log \left( \frac{\rho_\mathrm{sat}}{\rho_\mathrm{tot} - \rho_\mathrm{sat}} \right).\label{eq:porous-sat_cond}
\end{alignat}
Below, this result is used to relate the liquid free energy \(f_\mathrm{liq}\) to the vapor saturation pressure. Note that it is independent of the substrate state.

Similarly, equating the film and substrate pressures, \(\delta \mathcal{F} / \delta \psi_1 = \delta \mathcal{F} / \delta \psi_2\), allows one to define the equilibrium filling ratio for a substrate that is covered with a thick film
\begin{equation}
    \partial_\zeta \left[ \gamma_\mathrm{sl}(\Theta) + f_\text{wet}(h, \Theta) + z_0 \Psi(\Theta) \right] = 0.
\end{equation}
Next, equating the substrate and vapor pressures, \(\delta \mathcal{F} / \delta \psi_2 = \delta \mathcal{F} / \delta \psi_3\), gives the equilibrium substrate filling ratio as a function of the ambient vapor density.
\begin{equation}
    \partial_\zeta \left[ \gamma_\mathrm{sl}(\Theta) + f_\text{wet}(h_p,\,\Theta) + z_0 \Psi(\Theta) \right] + f_\text{liq} = \rho_\mathrm{liq} k_B T \log \left( \frac{\rho_\mathrm{vap}}{\rho_\mathrm{tot} - \rho_\mathrm{vap}} \right).
\end{equation}
Using Eq.~\eqref{eq:porous-sat_cond} to express the liquid bulk free energy through the saturation pressure, the equation can be written as
\begin{equation}
    \phi(\Theta) = \exp \left\{ \frac{\partial_\zeta \left[ \gamma_\mathrm{sl}(\Theta) + f_\text{wet}(h_p,\,\Theta) + z_0 \Psi(\Theta) \right] }{\rho_\mathrm{liq}k_BT} \right\}.\label{eq:porous-iso1}
\end{equation}
The result implicitly defines the sorption isotherm \(\Theta(\phi)\), i.e., the substrate state as a function of the vapor concentration.
If the porosity is low (or the substrate is sufficiently thick), the term related to the (bulk) capillary potential dominates the interface terms. This is implied by their scaling behavior with respect to the substrate thickness \(z_0\) or the porosity \(b\).  Thus, we simplify the definition of the isotherm by neglecting the interface terms and write
\begin{equation}
    \phi(\Theta) = \exp \left\{ \frac{\partial_\Theta \Psi(\Theta)}{b \rho_\mathrm{liq} k_B T} \right\}.\label{eq:porous-iso2}
\end{equation}
\begin{figure}[tb]
    \centering
    \includegraphics[width=0.8\hsize]{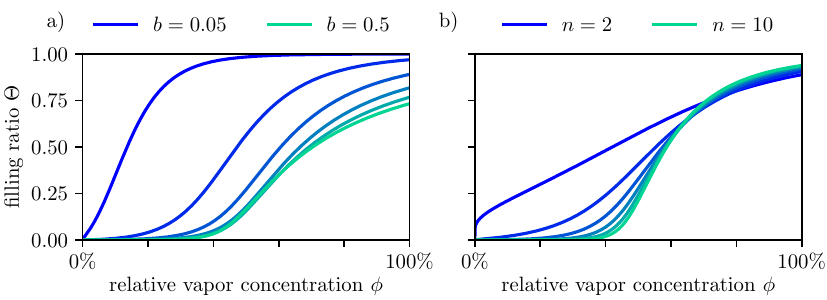}
    \caption{Sorption isotherms \(\Theta(\phi)\) of a porous medium in vapor according to Eq.~\eqref{eq:porous-iso2} for various parameter configurations and fixed dimensionless capillary pressure constant \(p_c / (\rho_\mathrm{liq} k_BT) = \num{0.1}\). In panel (a) we linearly vary the porosity \(b\) at fixed \(n=5\) and panel (b) shows the linear variation of the exponent \(n\) at fixed porosity \(b=\num{0.2}\). Note that due to entropic influences some air remains in the pores even for a saturated gas phase, \(\Theta(\phi=\SI{100}{\percent})<1\).}\label{fig:porous-isotherm}
\end{figure}
Fig.~\ref{fig:porous-isotherm} shows sorption isotherms \(\Theta(\phi)\) of a porous substrate at fixed dimensionless capillary pressure constant \(p_c / (\rho_\mathrm{liq} k_BT) = \num{0.1}\) for various values of the porosity \(b\) and the exponent \(n\).

\section{Spreading Dynamics: Halo Formation}
\label{sec:spreading}

\begin{table*}[!htb]
    \begin{tabular}{lcc}
        \textbf{Parameter name}                           & ~\textbf{Symbol}~        & \textbf{Value}\\
        \hline
viscosity                                         & \(\eta\)                 & \SI{0.89}{mPa\,s}\\
        contact angle parameter (dry substrate)           & \(\theta_\mathrm{eq}\)             & \(\tan \SI{30}{\degree}\)\\
        precursor layer height                            & \(h_p\)                  & \SI{10}{\micro m}\\
        liquid particle density                           & \(\rho_\mathrm{liq}\)    & \(\frac{\SI{997}{kg / m^3}}{\SI{18}{g / mol}} N_A\)\\
        vapor saturation pressure                         & \(p_\mathrm{sat}\)       & \SI{2643}{Pa}\\
        temperature                                       & \(T\)                    & \SI{22}{\celsius}\\
        initial drop volume                               & \(V_0\)                  & \SI{23}{\micro l}\\
        initial vapor concentration                       & \(\phi_\mathrm{lab}\)    & \SI{10}{\percent}\\
liquid-gas interface energy                       & \(\gamma_\mathrm{lg}\)   & \SI{72.8}{\milli N / m}\\
        substrate-liquid interface energy (dry substrate) & \(\gamma_\mathrm{bl,0}\) & \SI{3}{\milli N / m}\\
        imbibition front interface energy                 & \(\Gamma\)               & \SI{0.3}{\milli N / m}\\
substrate porosity                                & \(b\)                    & \SI{0.2}{}\\
        substrate thickness                               & \(z_0\)                  & \SI{10}{\micro m}\\
        pore size distribution coefficient                & \(n\)                    & 5\\
        adaption exponent (power law)                     & \(a\)                    & 1\\
        permeability                                      & \(\kappa\)               & \(\SI{2e-17}{m^2}\)\\
vapor diffusion coefficient                       & \(D_\mathrm{vap}\)       & \SI{2.82e-5}{m^2 / s}\\
        imbibition rate coefficient                       & \(M_\mathrm{im}\)        & \SI{e-13}{m / Pa\,s}\\
        bulk liquid evaporation rate coefficient          & \(M_\mathrm{ev}\)        & \SI{e-13}{m / Pa\,s}\\
        substrate evaporation rate coefficient            & \(M_\mathrm{ev}'\)       & \SI{e-13}{m / Pa\,s}\\
simulation domain height                          & \(d\)                    & \SI{3}{mm}\\
        simulation domain width                           & \(L\)                    & \SI{20}{mm}
    \end{tabular}
    \caption{Model parameters used for the numerical simulation of the coupled spreading, evaporation and imbibition dynamics in Fig.~\ref{fig:porous-snapshots}.}
    \label{tab:porous-params}
\end{table*}

\begin{figure}[!bt]
    \centering
    \includegraphics[width=0.8\hsize]{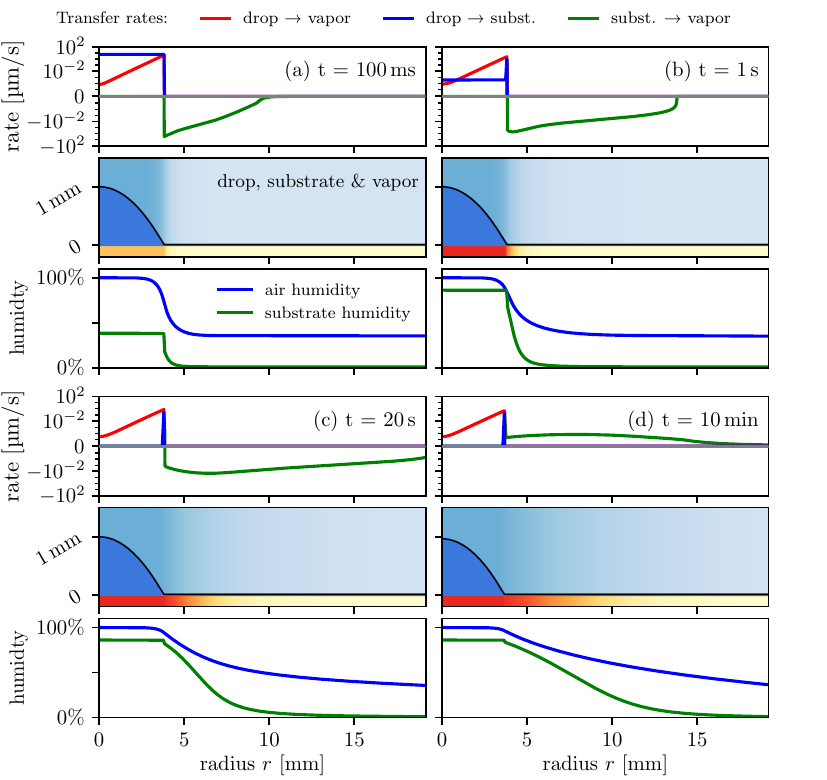}
    \caption{Snapshots at four selected times from a numerical simulation of the coupled spreading, evaporation and imbibition dynamics of a sessile drop on a porous substrate at parameters given in Tab.~\ref{tab:porous-params}. At each point in time (a)--(d) three subpanels are given: (top)  local transfer rates \(j_\mathrm{ev}\), \(j_\mathrm{ev}\)', and \(j_\mathrm{im}\) on a symmetric log-scale with a linear scale between \(\pm\SI{e-5}{\micro m/s}\) (volume per area per time); (center) simulation domain with the drop in dark blue, the vapor concentration \(\phi\) as blue shading, and the porous substrate filling state \(\Theta \) as red-to-yellow shading; (bottom) relative vapor and substrate humidities, \(\phi \) and \(\Theta \). For chosen times and various line styles see legends.}\label{fig:porous-snapshots}
\end{figure}

Next we consider an example for the coupled spreading, evaporation and imbibition dynamics of a sessile drop on a porous substrate.  Fig.~\ref{fig:porous-snapshots} shows four corresponding snapshots of a simulation of the developed three-field porous substrate model, Eq.~\eqref{eq:porous-model}, in a radial geometry. The boundary conditions are chosen such that the vapor particles are allowed to freely escape from the radial domain at the outer boundary (at \(r=\SI{20}{mm}\)), where the lab humidity \(\phi_\mathrm{lab} = \SI{10}{\percent}\) is enforced. The full set of parameters is given in Tab.~\ref{tab:porous-params}. The parameters represent realistic values, however, our choice is not geared towards a particular experiment. The simulation is performed employing the finite element method implemented in the package \texttt{oomph-lib}~\cite{HeHa2006}. Details regarding the model transformed into radial geometry can be found in Ref.~\citenum{Hartmann2023Munster}. 

At each of the four chosen times, Fig.~\ref{fig:porous-snapshots} contains three panels: The top panel shows the three transfer fluxes between the phases, namely, liquid-to-vapor evaporation/condensation flux \(j_\mathrm{ev}\), substrate-to-vapor evaporation/condensation flux \(j_\mathrm{ev}'\), and liquid-to-substrate imbibition flux \(j_\mathrm{im}\). The central panel visualizes the state of the system with the drop given in dark blue, the ambient vapor concentration presented as a light blue shading, and the porous substrate state as a red-to-yellow shade below the droplet. The shading represents the local filling ratio, where darker colors correspond to a larger liquid content of the substrate. The  bottom panel gives the humidity and filling ratio profiles of gas layer and porous substrate layer, respectively.

Inspection of Fig.~\ref{fig:porous-snapshots} indicates that in the initial second, fast imbibition of liquid occurs underneath the drop (green line in bottom panels of Fig.~\ref{fig:porous-snapshots}(a) and (b)). In parallel, evaporation results in saturation of the vapor phase directly above the droplet that here has the properties of water (red line in bottom panels of Fig.~\ref{fig:porous-snapshots}(a) and (b)). Interestingly, there is some absorption of liquid from the vapor into the porous layer outside the drop (green line in top panels of Fig.~\ref{fig:porous-snapshots}(a) and (b)).
Over the next minutes (Fig.~\ref{fig:porous-snapshots}(c) and (d)), lateral diffusion becomes the dominating process. On the one hand, the vapor diffuses what represent the limiting process for the evaporation of the drop as we are in the diffusion-limited regime, see Ref.~\cite{HDJT2023jfm} for an extensive discussion of the different regimes. On the other hand, also the substrate filling is extended far beyond the contact line of the drop. This occurs not only due to diffusion within the substrate but also because there is a significant condensation of liquid from the vapor phase into the substrate. We emphasize that the strength of this effect depends on the ratio between the various rates for the competing diffusive processes. For instance, if the substrate had a larger permeability, transport within the substrate could be faster than transport in the gas phase implying that one could in the first minutes observe evaporation of liquid from the substrate outside the drop into the gas phase.

Here, the condensation continues for about a minute (between Fig.~\ref{fig:porous-snapshots}(c) and (d)). Then, the process reverses and liquid evaporates from the porous layer into the gas phase. This results in the emergence of a quasi-stationary macroscopic ``halo'' in the substrate filling ratio. The system establishes a continuous flow of liquid from the drop radially outwards through the porous substrate. As it then evaporates thereby accelerating the transport through the gas phase the halo develops.
The resulting halo persists for the full time until the droplet has completely evaporated. Similar results have been obtained for the case of a liquid drop evaporating on a polymer-brush covered substrate~\cite{KHHB2023jcp}. However, there the brush does not only adapt its ``filling ratio'' but also its surface topography.

\section{Conclusion}
\label{sec:conc}

We have presented a mesoscopic hydrodynamic model in gradient dynamics formulation for a sessile drop of volatile partially wetting liquid on a solid porous layer that is able to describe situations where several dynamic processes compete that are all driven by the same underlying energy functional. The latter incorporates saturation-dependent wettability, capillarity of the drop and within the porous layer, air and vapor entropy and an entropic contribution relevant close to complete filling of the porous layer that avoids unphysical behavior. The modeled transport and transfer processes are the spreading of the droplet, the evaporation of liquid from the drop into the ambient gas or condensation of vapor from the ambient gas, vapor diffusion in the gas phase, the vertical absorption of liquid from the drop or the vapor into the porous substrate, and the lateral imbibition dynamics of the liquid within the porous substrate layer. These coupled processes are captured by spatially two-dimensional partial differential equations for the particle densities per substrate area in the three regions: liquid drop, porous layer and gas phase. These particle densities are closely related to the drop height profile, the vertically averaged saturation profile in the porous layer, and the vertically averaged vapor density profile, respectively. The approach assumes strong confinement, i.e., we have (i) considered a sessile drop on one of two parallel plates that form a narrow gap, and (ii) a thin porous substrate layer. This has generalized the approach of Ref.~\cite{HDJT2023jfm} for drops of volatile liquids within a gap between smooth solid plates, and partially parallels such models for drops on polymer brushes~\cite{HDGT2024l}. In contrast, here, the imbibition process is incorporated adapting and extending models for the dynamics of the saturation profile in a porous medium that represent diffuse-interface generalizations of Darcy's law and Richards' equation used, e.g., to study gravity-driven saturation fronts within porous media~\cite{CuJu2009wrr,CuJu2009pre,BeCa2022el}.

The incorporation of drop and vapor dynamics now allows for the adaptation of the extended model to a number of practical problems. An example important in the context of rhizospheric soil in the presence of plant root-produced exopolysaccharides is the water transfer between a drop and an amphiphilic porous medium, studied for nonvolatile liquids in~\cite{Cajot2024Avignon,CaDB2024preprint}. There, one could introduce the interplay between ambient humidity and saturation dynamics. Other examples that one could now investigate with consideration of both, imbibition and vapor dynamics, include inkjet printing on layers of paper~\cite{CBCW2002l} and on microporous membranes~\cite{BNSI2021csaea}, and sliding drops on porous substrates~\cite{GMBB2020pccp}. Furthermore, the approach could be expanded to cover various suspensions and solutions on porous substrates~\cite{Star2004acis,HCUS2002jcis} by combining the gradient dynamics approach presented here with elements of Refs.~\cite{ThTL2013prl,ThAP2016prf}.

In the present work we have given the mesoscopic hydrodynamic model in long-wave approximation~\cite{OrDB1997rmp}. However, as discussed in Ref.~\cite{Thie2018csa} in the spirit of the Cahn-Hilliard equation for phase separation one may also use a hybrid model where the mobilities are kept in long-wave approximation but the energy functional is made more precise by using a full-curvature formulation. It has been shown that this often results in more exact models, see e.g., Refs.~\cite{BoTH2018jfm,HDGT2024l,Diekmann2022Munster,DiTh2024preprint}. Here, the full-curvature formulation is obtained by replacing the long-wave metric factor $\frac{1}{2}(\nabla h)^2$ in Eq.~\eqref{eq:porous-energy} by the full expression $\sqrt{1+(\nabla h)^2}$. Note that the basic gradient dynamics can, in principle, also be adapted to situations like substrates with chemical or topographic heterogeneities~\cite{TBBB2003epje,SaKa2012jem}. The employed gradient dynamics approach also provides a relatively simple, but by definition thermodynamically consistent approach to more complex situations~\cite{ThAP2016prf,Thie2018csa,HEHZ2022prsa,HDGT2024l,DiTh2024epjst}. In this way, our present approach could in the future be further expanded to cover situations where the underlying governing equations are not yet completely known. Examples include systems where imbibition results in the swelling of the porous medium as for thin deformable porous media~\cite{KMHB2017prf}, liquid-infused slippery substrates that are often based on porous substrate layers~\cite{GWXM2015l,MMTM2016an,LGGM2017pra,BMWB2017apl}, imbibition into loose material as a powder bed~\cite{HLBH2002jcis}, spreading and imbibition of drops of liquid mixtures, and photocatalytically active lubricant-impregnated surfaces~\cite{WoBu2017acie}.

\section*{Data availability statement}
The underlying data needed to reproduce shown results are publicly available at the data repository \textit{zenodo}, see:\\
Hartmann, S. and Thiele, U. (2024). Data Supplement for: ``Gradient dynamics model for drops of volatile liquid on a porous substrate'' \textit{[Data set]. Zenodo.} \href{https://doi.org/10.5281/zenodo.13127822}{10.5281/zenodo.13127822}.

\acknowledgments

The authors acknowledge support by the Deutsche Forschungsgemeinschaft (DFG) via grant No.\ TH781/12-1 and TH781/12-2 within Priority Program (SPP)~2171 \enquote{Dynamic Wetting of Flexible, Adaptive, and Switchable Surfaces}; UT would further like to thank the Isaac Newton Institute for Mathematical Sciences, Cambridge, for support and hospitality during the program \textit{Anti-Diffusion in Multiphase and Active Flows (ADIW04)} where some work on this paper was undertaken. The program was supported by EPSRC grant No. EP/R014604/1. We acknowledge fruitful discussions with Florian Cajot, Philippe Beltrame, Laura Gallardo Dominguez, Patrick Huber as well as many discussions on adaptive substrates with Jan Diekmann, Daniel Greve, and Christopher Henkel at the University of M\"unster, and with participants of the events organized by SPP~2171.

\bibliography{literature}

\begin{thebibliography}{91}%
\makeatletter
\providecommand \@ifxundefined [1]{%
 \@ifx{#1\undefined}
}%
\providecommand \@ifnum [1]{%
 \ifnum #1\expandafter \@firstoftwo
 \else \expandafter \@secondoftwo
 \fi
}%
\providecommand \@ifx [1]{%
 \ifx #1\expandafter \@firstoftwo
 \else \expandafter \@secondoftwo
 \fi
}%
\providecommand \natexlab [1]{#1}%
\providecommand \enquote  [1]{``#1''}%
\providecommand \bibnamefont  [1]{#1}%
\providecommand \bibfnamefont [1]{#1}%
\providecommand \citenamefont [1]{#1}%
\providecommand \href@noop [0]{\@secondoftwo}%
\providecommand \href [0]{\begingroup \@sanitize@url \@href}%
\providecommand \@href[1]{\@@startlink{#1}\@@href}%
\providecommand \@@href[1]{\endgroup#1\@@endlink}%
\providecommand \@sanitize@url [0]{\catcode `\\12\catcode `\$12\catcode
  `\&12\catcode `\#12\catcode `\^12\catcode `\_12\catcode `\%12\relax}%
\providecommand \@@startlink[1]{}%
\providecommand \@@endlink[0]{}%
\providecommand \url  [0]{\begingroup\@sanitize@url \@url }%
\providecommand \@url [1]{\endgroup\@href {#1}{\urlprefix }}%
\providecommand \urlprefix  [0]{URL }%
\providecommand \Eprint [0]{\href }%
\providecommand \doibase [0]{https://doi.org/}%
\providecommand \selectlanguage [0]{\@gobble}%
\providecommand \bibinfo  [0]{\@secondoftwo}%
\providecommand \bibfield  [0]{\@secondoftwo}%
\providecommand \translation [1]{[#1]}%
\providecommand \BibitemOpen [0]{}%
\providecommand \bibitemStop [0]{}%
\providecommand \bibitemNoStop [0]{.\EOS\space}%
\providecommand \EOS [0]{\spacefactor3000\relax}%
\providecommand \BibitemShut  [1]{\csname bibitem#1\endcsname}%
\let\auto@bib@innerbib\@empty
%</preamble>
\bibitem [{\citenamefont {Richards}(1931)}]{Rich1931p}%
  \BibitemOpen
  \bibfield  {author} {\bibinfo {author} {\bibfnamefont {L.~A.}\ \bibnamefont
  {Richards}},\ }\bibfield  {title} {\bibinfo {title} {Capillary conduction of
  liquids through porous mediums},\ }\href {https://doi.org/10.1063/1.1745010}
  {\bibfield  {journal} {\bibinfo  {journal} {Physics}\ }\textbf {\bibinfo
  {volume} {1}},\ \bibinfo {pages} {318} (\bibinfo {year} {1931})}\BibitemShut
  {NoStop}%
\bibitem [{\citenamefont {Saffman}\ and\ \citenamefont
  {Taylor}(1958)}]{SaTa1958prslsa}%
  \BibitemOpen
  \bibfield  {author} {\bibinfo {author} {\bibfnamefont {P.~G.}\ \bibnamefont
  {Saffman}}\ and\ \bibinfo {author} {\bibfnamefont {G.}~\bibnamefont
  {Taylor}},\ }\bibfield  {title} {\bibinfo {title} {The penetration of a fluid
  into a porous medium or {H}ele-{S}haw cell containing a more viscous
  liquid},\ }\href {https://doi.org/10.1016/b978-0-08-092523-3.50017-4}
  {\bibfield  {journal} {\bibinfo  {journal} {Proc. R. Soc. London Ser. A}\
  }\textbf {\bibinfo {volume} {245}},\ \bibinfo {pages} {312} (\bibinfo {year}
  {1958})}\BibitemShut {NoStop}%
\bibitem [{\citenamefont {Beavers}\ and\ \citenamefont
  {Joseph}(1967)}]{BeJo1967jfm}%
  \BibitemOpen
  \bibfield  {author} {\bibinfo {author} {\bibfnamefont {G.~S.}\ \bibnamefont
  {Beavers}}\ and\ \bibinfo {author} {\bibfnamefont {D.~D.}\ \bibnamefont
  {Joseph}},\ }\bibfield  {title} {\bibinfo {title} {Boundary conditions at a
  naturally permeable wall},\ }\href
  {https://doi.org/10.1017/s0022112067001375} {\bibfield  {journal} {\bibinfo
  {journal} {J. Fluid Mech.}\ }\textbf {\bibinfo {volume} {30}},\ \bibinfo
  {pages} {197} (\bibinfo {year} {1967})}\BibitemShut {NoStop}%
\bibitem [{\citenamefont {Whitaker}(1986)}]{Whit1986tpm}%
  \BibitemOpen
  \bibfield  {author} {\bibinfo {author} {\bibfnamefont {S.}~\bibnamefont
  {Whitaker}},\ }\bibfield  {title} {\bibinfo {title} {Flow in porous media
  {I}: A theoretical derivation of {D}arcy's law},\ }\href
  {https://doi.org/10.1007/BF01036523} {\bibfield  {journal} {\bibinfo
  {journal} {Transp. Porous Media}\ }\textbf {\bibinfo {volume} {1}},\ \bibinfo
  {pages} {3} (\bibinfo {year} {1986})}\BibitemShut {NoStop}%
\bibitem [{\citenamefont {Marmur}(1988)}]{Marm1988jcis}%
  \BibitemOpen
  \bibfield  {author} {\bibinfo {author} {\bibfnamefont {A.}~\bibnamefont
  {Marmur}},\ }\bibfield  {title} {\bibinfo {title} {Drop penetration into a
  thin porous-medium},\ }\href {https://doi.org/10.1016/0021-9797(88)90233-0}
  {\bibfield  {journal} {\bibinfo  {journal} {J. Colloid Interface Sci.}\
  }\textbf {\bibinfo {volume} {123}},\ \bibinfo {pages} {161} (\bibinfo {year}
  {1988})}\BibitemShut {NoStop}%
\bibitem [{\citenamefont {Davis}\ and\ \citenamefont
  {Hocking}(1999)}]{DaHo1999pf}%
  \BibitemOpen
  \bibfield  {author} {\bibinfo {author} {\bibfnamefont {S.~H.}\ \bibnamefont
  {Davis}}\ and\ \bibinfo {author} {\bibfnamefont {L.~M.}\ \bibnamefont
  {Hocking}},\ }\bibfield  {title} {\bibinfo {title} {Spreading and imbibition
  of viscous liquid on a porous base},\ }\href
  {https://doi.org/10.1063/1.869901} {\bibfield  {journal} {\bibinfo  {journal}
  {Phys. Fluids}\ }\textbf {\bibinfo {volume} {11}},\ \bibinfo {pages} {48}
  (\bibinfo {year} {1999})}\BibitemShut {NoStop}%
\bibitem [{\citenamefont {Vasin}\ \emph {et~al.}(1996)\citenamefont {Vasin},
  \citenamefont {Starov},\ and\ \citenamefont {Filippov}}]{VaSF1996cj}%
  \BibitemOpen
  \bibfield  {author} {\bibinfo {author} {\bibfnamefont {S.~I.}\ \bibnamefont
  {Vasin}}, \bibinfo {author} {\bibfnamefont {V.~M.}\ \bibnamefont {Starov}},\
  and\ \bibinfo {author} {\bibfnamefont {A.~N.}\ \bibnamefont {Filippov}},\
  }\bibfield  {title} {\bibinfo {title} {The motion of a solid spherical
  particle covered with a porous layer in a liquid},\ }\href@noop {} {\bibfield
   {journal} {\bibinfo  {journal} {Colloid J.}\ }\textbf {\bibinfo {volume}
  {58}},\ \bibinfo {pages} {282} (\bibinfo {year} {1996})}\BibitemShut
  {NoStop}%
\bibitem [{\citenamefont {Worster}(1992)}]{Wors1992jfm}%
  \BibitemOpen
  \bibfield  {author} {\bibinfo {author} {\bibfnamefont {M.~G.}\ \bibnamefont
  {Worster}},\ }\bibfield  {title} {\bibinfo {title} {Instabilities of the
  liquid and mushy regions during solidification of alloys},\ }\href
  {https://doi.org/10.1017/S0022112092003562} {\bibfield  {journal} {\bibinfo
  {journal} {J. Fluid Mech.}\ }\textbf {\bibinfo {volume} {237}},\ \bibinfo
  {pages} {649} (\bibinfo {year} {1992})}\BibitemShut {NoStop}%
\bibitem [{\citenamefont {Davis}\ and\ \citenamefont
  {Hocking}(2000)}]{DaHo2000pf}%
  \BibitemOpen
  \bibfield  {author} {\bibinfo {author} {\bibfnamefont {S.~H.}\ \bibnamefont
  {Davis}}\ and\ \bibinfo {author} {\bibfnamefont {L.~M.}\ \bibnamefont
  {Hocking}},\ }\bibfield  {title} {\bibinfo {title} {Spreading and imbibition
  of viscous liquid on a porous base. {II}},\ }\href
  {https://doi.org/10.1063/1.870416} {\bibfield  {journal} {\bibinfo  {journal}
  {Phys. Fluids}\ }\textbf {\bibinfo {volume} {12}},\ \bibinfo {pages} {1646}
  (\bibinfo {year} {2000})}\BibitemShut {NoStop}%
\bibitem [{\citenamefont {Starov}\ \emph {et~al.}(2003)\citenamefont {Starov},
  \citenamefont {Zhdanov}, \citenamefont {Kosvintsev}, \citenamefont
  {Sobolev},\ and\ \citenamefont {Velarde}}]{SZKS2003acis}%
  \BibitemOpen
  \bibfield  {author} {\bibinfo {author} {\bibfnamefont {V.~M.}\ \bibnamefont
  {Starov}}, \bibinfo {author} {\bibfnamefont {S.~A.}\ \bibnamefont {Zhdanov}},
  \bibinfo {author} {\bibfnamefont {S.~R.}\ \bibnamefont {Kosvintsev}},
  \bibinfo {author} {\bibfnamefont {V.~D.}\ \bibnamefont {Sobolev}},\ and\
  \bibinfo {author} {\bibfnamefont {M.~G.}\ \bibnamefont {Velarde}},\
  }\bibfield  {title} {\bibinfo {title} {Spreading of liquid drops over porous
  substrates},\ }\href {https://doi.org/10.1016/s0001-8686(03)00039-3}
  {\bibfield  {journal} {\bibinfo  {journal} {Adv. Colloid Interface Sci.}\
  }\textbf {\bibinfo {volume} {104}},\ \bibinfo {pages} {123} (\bibinfo {year}
  {2003})}\BibitemShut {NoStop}%
\bibitem [{\citenamefont {Alleborn}\ and\ \citenamefont
  {Raszillier}(2004)}]{AlRa2004ces}%
  \BibitemOpen
  \bibfield  {author} {\bibinfo {author} {\bibfnamefont {N.}~\bibnamefont
  {Alleborn}}\ and\ \bibinfo {author} {\bibfnamefont {H.}~\bibnamefont
  {Raszillier}},\ }\bibfield  {title} {\bibinfo {title} {Spreading and sorption
  of a droplet on a porous substrate},\ }\href
  {https://doi.org/10.1016/j.ces.2004.02.006} {\bibfield  {journal} {\bibinfo
  {journal} {Chem. Eng. Sci.}\ }\textbf {\bibinfo {volume} {59}},\ \bibinfo
  {pages} {2071} (\bibinfo {year} {2004})}\BibitemShut {NoStop}%
\bibitem [{\citenamefont {Xiao}\ \emph {et~al.}(2012)\citenamefont {Xiao},
  \citenamefont {Stone},\ and\ \citenamefont {Attinger}}]{XiSA2012l}%
  \BibitemOpen
  \bibfield  {author} {\bibinfo {author} {\bibfnamefont {J.~F.}\ \bibnamefont
  {Xiao}}, \bibinfo {author} {\bibfnamefont {H.~A.}\ \bibnamefont {Stone}},\
  and\ \bibinfo {author} {\bibfnamefont {D.}~\bibnamefont {Attinger}},\
  }\bibfield  {title} {\bibinfo {title} {Source-like solution for radial
  imbibition into a homogeneous semi-infinite porous medium},\ }\href
  {https://doi.org/10.1021/la204474f} {\bibfield  {journal} {\bibinfo
  {journal} {Langmuir}\ }\textbf {\bibinfo {volume} {28}},\ \bibinfo {pages}
  {4208} (\bibinfo {year} {2012})}\BibitemShut {NoStop}%
\bibitem [{\citenamefont {Gambaryan-Roisman}(2014)}]{Gamb2014cocis}%
  \BibitemOpen
  \bibfield  {author} {\bibinfo {author} {\bibfnamefont {T.}~\bibnamefont
  {Gambaryan-Roisman}},\ }\bibfield  {title} {\bibinfo {title} {Liquids on
  porous layers: wetting, imbibition and transport processes},\ }\href
  {https://doi.org/10.1016/j.cocis.2014.09.001} {\bibfield  {journal} {\bibinfo
   {journal} {Curr. Opin. Colloid Interface Sci.}\ }\textbf {\bibinfo {volume}
  {19}},\ \bibinfo {pages} {320} (\bibinfo {year} {2014})}\BibitemShut
  {NoStop}%
\bibitem [{\citenamefont {Arjmandi-Tash}\ \emph {et~al.}(2017)\citenamefont
  {Arjmandi-Tash}, \citenamefont {Koyalchuk}, \citenamefont {Trybala},
  \citenamefont {Kuchin},\ and\ \citenamefont {Starov}}]{AKTK2017l}%
  \BibitemOpen
  \bibfield  {author} {\bibinfo {author} {\bibfnamefont {O.}~\bibnamefont
  {Arjmandi-Tash}}, \bibinfo {author} {\bibfnamefont {N.~M.}\ \bibnamefont
  {Koyalchuk}}, \bibinfo {author} {\bibfnamefont {A.}~\bibnamefont {Trybala}},
  \bibinfo {author} {\bibfnamefont {I.~V.}\ \bibnamefont {Kuchin}},\ and\
  \bibinfo {author} {\bibfnamefont {V.}~\bibnamefont {Starov}},\ }\bibfield
  {title} {\bibinfo {title} {Kinetics of wetting and spreading of droplets over
  various substrates},\ }\href {https://doi.org/10.1021/acs.langmuir.6b04094}
  {\bibfield  {journal} {\bibinfo  {journal} {Langmuir}\ }\textbf {\bibinfo
  {volume} {33}},\ \bibinfo {pages} {4367} (\bibinfo {year}
  {2017})}\BibitemShut {NoStop}%
\bibitem [{\citenamefont {Yao}\ \emph {et~al.}(2017)\citenamefont {Yao},
  \citenamefont {Alexandris}, \citenamefont {Henrich}, \citenamefont
  {Auernhammer}, \citenamefont {Steinhart}, \citenamefont {Butt},\ and\
  \citenamefont {Floudas}}]{YAHA2017jcp}%
  \BibitemOpen
  \bibfield  {author} {\bibinfo {author} {\bibfnamefont {Y.}~\bibnamefont
  {Yao}}, \bibinfo {author} {\bibfnamefont {S.}~\bibnamefont {Alexandris}},
  \bibinfo {author} {\bibfnamefont {F.}~\bibnamefont {Henrich}}, \bibinfo
  {author} {\bibfnamefont {G.}~\bibnamefont {Auernhammer}}, \bibinfo {author}
  {\bibfnamefont {M.}~\bibnamefont {Steinhart}}, \bibinfo {author}
  {\bibfnamefont {H.~J.}\ \bibnamefont {Butt}},\ and\ \bibinfo {author}
  {\bibfnamefont {G.}~\bibnamefont {Floudas}},\ }\bibfield  {title} {\bibinfo
  {title} {Complex dynamics of capillary imbibition of poly(ethylene oxide)
  melts in nanoporous alumina},\ }\href {https://doi.org/10.1063/1.4978298}
  {\bibfield  {journal} {\bibinfo  {journal} {J. Chem. Phys.}\ }\textbf
  {\bibinfo {volume} {146}},\ \bibinfo {pages} {203320} (\bibinfo {year}
  {2017})}\BibitemShut {NoStop}%
\bibitem [{\citenamefont {Johnson}\ \emph {et~al.}(2019)\citenamefont
  {Johnson}, \citenamefont {Trybala},\ and\ \citenamefont
  {Starov}}]{JoTS2019ci}%
  \BibitemOpen
  \bibfield  {author} {\bibinfo {author} {\bibfnamefont {P.}~\bibnamefont
  {Johnson}}, \bibinfo {author} {\bibfnamefont {A.}~\bibnamefont {Trybala}},\
  and\ \bibinfo {author} {\bibfnamefont {V.}~\bibnamefont {Starov}},\
  }\bibfield  {title} {\bibinfo {title} {Kinetics of spreading over porous
  substrates},\ }\href {https://doi.org/10.3390/colloids3010038} {\bibfield
  {journal} {\bibinfo  {journal} {Colloid Interfac.}\ }\textbf {\bibinfo
  {volume} {3}},\ \bibinfo {pages} {38} (\bibinfo {year} {2019})}\BibitemShut
  {NoStop}%
\bibitem [{\citenamefont {Etha}\ \emph {et~al.}(2021)\citenamefont {Etha},
  \citenamefont {Desai}, \citenamefont {Sachar},\ and\ \citenamefont
  {Das}}]{EDSD2021m}%
  \BibitemOpen
  \bibfield  {author} {\bibinfo {author} {\bibfnamefont {S.~A.}\ \bibnamefont
  {Etha}}, \bibinfo {author} {\bibfnamefont {P.~R.}\ \bibnamefont {Desai}},
  \bibinfo {author} {\bibfnamefont {H.~S.}\ \bibnamefont {Sachar}},\ and\
  \bibinfo {author} {\bibfnamefont {S.}~\bibnamefont {Das}},\ }\bibfield
  {title} {\bibinfo {title} {Wetting dynamics on solvophilic, soft, porous, and
  responsive surfaces},\ }\href {https://doi.org/10.1021/acs.macromol.0c02234}
  {\bibfield  {journal} {\bibinfo  {journal} {Macromolecules}\ }\textbf
  {\bibinfo {volume} {54}},\ \bibinfo {pages} {584} (\bibinfo {year}
  {2021})}\BibitemShut {NoStop}%
\bibitem [{\citenamefont {de~Gennes}(1999)}]{Genn1999crasi}%
  \BibitemOpen
  \bibfield  {author} {\bibinfo {author} {\bibfnamefont {P.-G.}\ \bibnamefont
  {de~Gennes}},\ }\bibfield  {title} {\bibinfo {title} {Running droplets in a
  random medium},\ }\href@noop {} {\bibfield  {journal} {\bibinfo  {journal}
  {C. R. Acad. Sci. II}\ }\textbf {\bibinfo {volume} {327}},\ \bibinfo {pages}
  {147} (\bibinfo {year} {1999})}\BibitemShut {NoStop}%
\bibitem [{\citenamefont {Aradian}\ \emph {et~al.}(2000)\citenamefont
  {Aradian}, \citenamefont {Rapha{\"e}l},\ and\ \citenamefont
  {de~Gennes}}]{ArRG2000epje}%
  \BibitemOpen
  \bibfield  {author} {\bibinfo {author} {\bibfnamefont {A.}~\bibnamefont
  {Aradian}}, \bibinfo {author} {\bibfnamefont {E.}~\bibnamefont
  {Rapha{\"e}l}},\ and\ \bibinfo {author} {\bibfnamefont {P.~G.}\ \bibnamefont
  {de~Gennes}},\ }\bibfield  {title} {\bibinfo {title} {Dewetting on porous
  media with aspiration},\ }\href {https://doi.org/10.1007/s101890050019}
  {\bibfield  {journal} {\bibinfo  {journal} {Eur. Phys. J. E}\ }\textbf
  {\bibinfo {volume} {2}},\ \bibinfo {pages} {367} (\bibinfo {year}
  {2000})}\BibitemShut {NoStop}%
\bibitem [{\citenamefont {Bacri}\ and\ \citenamefont
  {Brochard-Wyart}(2001)}]{BaBr2001el}%
  \BibitemOpen
  \bibfield  {author} {\bibinfo {author} {\bibfnamefont {L.}~\bibnamefont
  {Bacri}}\ and\ \bibinfo {author} {\bibfnamefont {F.}~\bibnamefont
  {Brochard-Wyart}},\ }\bibfield  {title} {\bibinfo {title} {Dewetting on
  porous media},\ }\href@noop {} {\bibfield  {journal} {\bibinfo  {journal}
  {Europhys. Lett.}\ }\textbf {\bibinfo {volume} {56}},\ \bibinfo {pages} {414}
  (\bibinfo {year} {2001})}\BibitemShut {NoStop}%
\bibitem [{\citenamefont {Desaive}\ \emph {et~al.}(2001)\citenamefont
  {Desaive}, \citenamefont {Lebon},\ and\ \citenamefont
  {Hennenberg}}]{DeLH2001pre}%
  \BibitemOpen
  \bibfield  {author} {\bibinfo {author} {\bibfnamefont {T.}~\bibnamefont
  {Desaive}}, \bibinfo {author} {\bibfnamefont {G.}~\bibnamefont {Lebon}},\
  and\ \bibinfo {author} {\bibfnamefont {M.}~\bibnamefont {Hennenberg}},\
  }\bibfield  {title} {\bibinfo {title} {Coupled capillary and gravity-driven
  instability in a liquid film overlying a porous layer},\ }\href@noop {}
  {\bibfield  {journal} {\bibinfo  {journal} {Phys. Rev. E}\ }\textbf {\bibinfo
  {volume} {64}},\ \bibinfo {pages} {066304} (\bibinfo {year}
  {2001})}\BibitemShut {NoStop}%
\bibitem [{\citenamefont {Sadiq}\ \emph {et~al.}(2010)\citenamefont {Sadiq},
  \citenamefont {Usha},\ and\ \citenamefont {Joo}}]{SaUJ2010ces}%
  \BibitemOpen
  \bibfield  {author} {\bibinfo {author} {\bibfnamefont {I.~M.~R.}\
  \bibnamefont {Sadiq}}, \bibinfo {author} {\bibfnamefont {R.}~\bibnamefont
  {Usha}},\ and\ \bibinfo {author} {\bibfnamefont {S.~W.}\ \bibnamefont
  {Joo}},\ }\bibfield  {title} {\bibinfo {title} {Instabilities in a liquid
  film flow over an inclined heated porous substrate},\ }\href
  {https://doi.org/10.1016/j.ces.2010.04.005} {\bibfield  {journal} {\bibinfo
  {journal} {Chem. Eng. Sci.}\ }\textbf {\bibinfo {volume} {65}},\ \bibinfo
  {pages} {4443} (\bibinfo {year} {2010})}\BibitemShut {NoStop}%
\bibitem [{\citenamefont {Cueto-Felgueroso}\ and\ \citenamefont
  {Juanes}(2008)}]{CuJu2008prl}%
  \BibitemOpen
  \bibfield  {author} {\bibinfo {author} {\bibfnamefont {L.}~\bibnamefont
  {Cueto-Felgueroso}}\ and\ \bibinfo {author} {\bibfnamefont {R.}~\bibnamefont
  {Juanes}},\ }\bibfield  {title} {\bibinfo {title} {Nonlocal interface
  dynamics and pattern formation in gravity-driven unsaturated flow through
  porous media},\ }\href {https://doi.org/10.1103/PhysRevLett.101.244504}
  {\bibfield  {journal} {\bibinfo  {journal} {Phys. Rev. Lett.}\ }\textbf
  {\bibinfo {volume} {101}},\ \bibinfo {pages} {244504} (\bibinfo {year}
  {2008})}\BibitemShut {NoStop}%
\bibitem [{\citenamefont {Cueto-Felgueroso}\ and\ \citenamefont
  {Juanes}(2009{\natexlab{a}})}]{CuJu2009wrr}%
  \BibitemOpen
  \bibfield  {author} {\bibinfo {author} {\bibfnamefont {L.}~\bibnamefont
  {Cueto-Felgueroso}}\ and\ \bibinfo {author} {\bibfnamefont {R.}~\bibnamefont
  {Juanes}},\ }\bibfield  {title} {\bibinfo {title} {A phase field model of
  unsaturated flow},\ }\href {https://doi.org/10.1029/2009WR007945} {\bibfield
  {journal} {\bibinfo  {journal} {Water Resour. Res.}\ }\textbf {\bibinfo
  {volume} {45}},\ \bibinfo {pages} {W10409} (\bibinfo {year}
  {2009}{\natexlab{a}})}\BibitemShut {NoStop}%
\bibitem [{\citenamefont {Cueto-Felgueroso}\ and\ \citenamefont
  {Juanes}(2009{\natexlab{b}})}]{CuJu2009pre}%
  \BibitemOpen
  \bibfield  {author} {\bibinfo {author} {\bibfnamefont {L.}~\bibnamefont
  {Cueto-Felgueroso}}\ and\ \bibinfo {author} {\bibfnamefont {R.}~\bibnamefont
  {Juanes}},\ }\bibfield  {title} {\bibinfo {title} {Stability analysis of a
  phase-field model of gravity-driven unsaturated flow through porous media},\
  }\href {https://doi.org/10.1103/PhysRevE.79.036301} {\bibfield  {journal}
  {\bibinfo  {journal} {Phys. Rev. E}\ }\textbf {\bibinfo {volume} {79}},\
  \bibinfo {pages} {036301} (\bibinfo {year} {2009}{\natexlab{b}})}\BibitemShut
  {NoStop}%
\bibitem [{\citenamefont {Devauchelle}\ \emph {et~al.}(2007)\citenamefont
  {Devauchelle}, \citenamefont {Josserand},\ and\ \citenamefont
  {Zaleski}}]{DeJZ2007jfm}%
  \BibitemOpen
  \bibfield  {author} {\bibinfo {author} {\bibfnamefont {O.}~\bibnamefont
  {Devauchelle}}, \bibinfo {author} {\bibfnamefont {C.}~\bibnamefont
  {Josserand}},\ and\ \bibinfo {author} {\bibfnamefont {S.}~\bibnamefont
  {Zaleski}},\ }\bibfield  {title} {\bibinfo {title} {Forced dewetting on
  porous media},\ }\href {https://doi.org/10.1017/S0022112006004125} {\bibfield
   {journal} {\bibinfo  {journal} {J. Fluid Mech.}\ }\textbf {\bibinfo {volume}
  {574}},\ \bibinfo {pages} {343} (\bibinfo {year} {2007})}\BibitemShut
  {NoStop}%
\bibitem [{\citenamefont {Goda}\ \emph {et~al.}(2017)\citenamefont {Goda},
  \citenamefont {Sasaki}, \citenamefont {Mizuno}, \citenamefont {Morizawa},
  \citenamefont {Katakura},\ and\ \citenamefont {Tomiya}}]{GSMM2017jctr}%
  \BibitemOpen
  \bibfield  {author} {\bibinfo {author} {\bibfnamefont {T.}~\bibnamefont
  {Goda}}, \bibinfo {author} {\bibfnamefont {Y.}~\bibnamefont {Sasaki}},
  \bibinfo {author} {\bibfnamefont {M.}~\bibnamefont {Mizuno}}, \bibinfo
  {author} {\bibfnamefont {K.}~\bibnamefont {Morizawa}}, \bibinfo {author}
  {\bibfnamefont {H.}~\bibnamefont {Katakura}},\ and\ \bibinfo {author}
  {\bibfnamefont {S.}~\bibnamefont {Tomiya}},\ }\bibfield  {title} {\bibinfo
  {title} {Numerical analysis for predicting the operability window of slot-die
  coating onto porous media},\ }\href
  {https://doi.org/10.1007/s11998-017-9985-7} {\bibfield  {journal} {\bibinfo
  {journal} {J. Coat. Technol. Res.}\ }\textbf {\bibinfo {volume} {14}},\
  \bibinfo {pages} {1053} (\bibinfo {year} {2017})}\BibitemShut {NoStop}%
\bibitem [{\citenamefont {Huber}(2015)}]{Hube2015jpcm}%
  \BibitemOpen
  \bibfield  {author} {\bibinfo {author} {\bibfnamefont {P.}~\bibnamefont
  {Huber}},\ }\bibfield  {title} {\bibinfo {title} {Soft matter in hard
  confinement: {P}hase transition thermodynamics, structure, texture, diffusion
  and flow in nanoporous media},\ }\href
  {https://doi.org/10.1088/0953-8984/27/10/103102} {\bibfield  {journal}
  {\bibinfo  {journal} {J. Phys.: Condens. Matter}\ }\textbf {\bibinfo {volume}
  {27}},\ \bibinfo {pages} {103102} (\bibinfo {year} {2015})}\BibitemShut
  {NoStop}%
\bibitem [{\citenamefont {Wang}\ \emph {et~al.}(2015)\citenamefont {Wang},
  \citenamefont {Heinke}, \citenamefont {Jelic}, \citenamefont {Cakici},
  \citenamefont {Dommaschk}, \citenamefont {Maurer}, \citenamefont {Oberhofer},
  \citenamefont {Grosjean}, \citenamefont {Herges}, \citenamefont {Brase},
  \citenamefont {Reuter},\ and\ \citenamefont {Woll}}]{WHJC2015pccp}%
  \BibitemOpen
  \bibfield  {author} {\bibinfo {author} {\bibfnamefont {Z.~B.}\ \bibnamefont
  {Wang}}, \bibinfo {author} {\bibfnamefont {L.}~\bibnamefont {Heinke}},
  \bibinfo {author} {\bibfnamefont {J.}~\bibnamefont {Jelic}}, \bibinfo
  {author} {\bibfnamefont {M.}~\bibnamefont {Cakici}}, \bibinfo {author}
  {\bibfnamefont {M.}~\bibnamefont {Dommaschk}}, \bibinfo {author}
  {\bibfnamefont {R.~J.}\ \bibnamefont {Maurer}}, \bibinfo {author}
  {\bibfnamefont {H.}~\bibnamefont {Oberhofer}}, \bibinfo {author}
  {\bibfnamefont {S.}~\bibnamefont {Grosjean}}, \bibinfo {author}
  {\bibfnamefont {R.}~\bibnamefont {Herges}}, \bibinfo {author} {\bibfnamefont
  {S.}~\bibnamefont {Brase}}, \bibinfo {author} {\bibfnamefont
  {K.}~\bibnamefont {Reuter}},\ and\ \bibinfo {author} {\bibfnamefont
  {C.}~\bibnamefont {Woll}},\ }\bibfield  {title} {\bibinfo {title}
  {Photoswitching in nanoporous, crystalline solids: an experimental and
  theoretical study for azobenzene linkers incorporated in mofs},\ }\href
  {https://doi.org/10.1039/c5cp01372k} {\bibfield  {journal} {\bibinfo
  {journal} {Phys. Chem. Chem. Phys.}\ }\textbf {\bibinfo {volume} {17}},\
  \bibinfo {pages} {14582} (\bibinfo {year} {2015})}\BibitemShut {NoStop}%
\bibitem [{\citenamefont {Bayada}\ and\ \citenamefont
  {Chambat}(1995)}]{BaCh1995sjma}%
  \BibitemOpen
  \bibfield  {author} {\bibinfo {author} {\bibfnamefont {G.}~\bibnamefont
  {Bayada}}\ and\ \bibinfo {author} {\bibfnamefont {M.}~\bibnamefont
  {Chambat}},\ }\bibfield  {title} {\bibinfo {title} {On interface conditions
  for a thin film flow past a porous medium},\ }\href
  {https://doi.org/10.1137/s0036141093246610} {\bibfield  {journal} {\bibinfo
  {journal} {SIAM J. Math. Anal.}\ }\textbf {\bibinfo {volume} {26}},\ \bibinfo
  {pages} {1113} (\bibinfo {year} {1995})}\BibitemShut {NoStop}%
\bibitem [{\citenamefont {Ochoa-Tapia}\ and\ \citenamefont
  {Whitaker}(1995{\natexlab{a}})}]{OcWh1995ijhmt}%
  \BibitemOpen
  \bibfield  {author} {\bibinfo {author} {\bibfnamefont {J.~A.}\ \bibnamefont
  {Ochoa-Tapia}}\ and\ \bibinfo {author} {\bibfnamefont {S.}~\bibnamefont
  {Whitaker}},\ }\bibfield  {title} {\bibinfo {title} {Momentum transfer at the
  boundary between a porous medium and a homogeneous fluid: {I. T}heoretical
  development},\ }\href {https://doi.org/10.1016/0017-9310(94)00346-W}
  {\bibfield  {journal} {\bibinfo  {journal} {Int. J. Heat Mass Transfer}\
  }\textbf {\bibinfo {volume} {38}},\ \bibinfo {pages} {2635} (\bibinfo {year}
  {1995}{\natexlab{a}})}\BibitemShut {NoStop}%
\bibitem [{\citenamefont {Ochoa-Tapia}\ and\ \citenamefont
  {Whitaker}(1995{\natexlab{b}})}]{OcWh1995ijhmtb}%
  \BibitemOpen
  \bibfield  {author} {\bibinfo {author} {\bibfnamefont {J.~A.}\ \bibnamefont
  {Ochoa-Tapia}}\ and\ \bibinfo {author} {\bibfnamefont {S.}~\bibnamefont
  {Whitaker}},\ }\bibfield  {title} {\bibinfo {title} {Momentum transfer at the
  boundary between a porous medium and a homogeneous fluid: {II. C}omparison
  with experiment},\ }\href {https://doi.org/10.1016/0017-9310(94)00347-X}
  {\bibfield  {journal} {\bibinfo  {journal} {Int. J. Heat Mass Transfer}\
  }\textbf {\bibinfo {volume} {38}},\ \bibinfo {pages} {2647} (\bibinfo {year}
  {1995}{\natexlab{b}})}\BibitemShut {NoStop}%
\bibitem [{\citenamefont {Jeong}(2001)}]{Jeon2001pf}%
  \BibitemOpen
  \bibfield  {author} {\bibinfo {author} {\bibfnamefont {J.~T.}\ \bibnamefont
  {Jeong}},\ }\bibfield  {title} {\bibinfo {title} {Slip boundary condition on
  an idealized porous wall},\ }\href {https://doi.org/10.1063/1.1373680}
  {\bibfield  {journal} {\bibinfo  {journal} {Phys. Fluids}\ }\textbf {\bibinfo
  {volume} {13}},\ \bibinfo {pages} {1884} (\bibinfo {year}
  {2001})}\BibitemShut {NoStop}%
\bibitem [{\citenamefont {Goyeau}\ \emph {et~al.}(2003)\citenamefont {Goyeau},
  \citenamefont {Lhuillier}, \citenamefont {Gobin},\ and\ \citenamefont
  {Velarde}}]{GLGV2003ijhmt}%
  \BibitemOpen
  \bibfield  {author} {\bibinfo {author} {\bibfnamefont {B.}~\bibnamefont
  {Goyeau}}, \bibinfo {author} {\bibfnamefont {D.}~\bibnamefont {Lhuillier}},
  \bibinfo {author} {\bibfnamefont {D.}~\bibnamefont {Gobin}},\ and\ \bibinfo
  {author} {\bibfnamefont {M.~G.}\ \bibnamefont {Velarde}},\ }\bibfield
  {title} {\bibinfo {title} {Momentum transport at a fluid-porous interface},\
  }\href {https://doi.org/10.1016/S0017-9310(03)00241-2} {\bibfield  {journal}
  {\bibinfo  {journal} {Int. J. Heat Mass Transf.}\ }\textbf {\bibinfo {volume}
  {46}},\ \bibinfo {pages} {4071} (\bibinfo {year} {2003})}\BibitemShut
  {NoStop}%
\bibitem [{\citenamefont {Valdes-Parada}\ \emph {et~al.}(2007)\citenamefont
  {Valdes-Parada}, \citenamefont {Goyeau},\ and\ \citenamefont
  {Ochoa-Tapia}}]{VaGO2007ces}%
  \BibitemOpen
  \bibfield  {author} {\bibinfo {author} {\bibfnamefont {F.~J.}\ \bibnamefont
  {Valdes-Parada}}, \bibinfo {author} {\bibfnamefont {B.}~\bibnamefont
  {Goyeau}},\ and\ \bibinfo {author} {\bibfnamefont {J.~A.}\ \bibnamefont
  {Ochoa-Tapia}},\ }\bibfield  {title} {\bibinfo {title} {Jump momentum
  boundary condition at a fluid-porous dividing surface: {D}erivation of the
  closure problem},\ }\href {https://doi.org/10.1016/j.ces.2007.04.042}
  {\bibfield  {journal} {\bibinfo  {journal} {Chem. Eng. Sci.}\ }\textbf
  {\bibinfo {volume} {62}},\ \bibinfo {pages} {4025} (\bibinfo {year}
  {2007})}\BibitemShut {NoStop}%
\bibitem [{\citenamefont {Thiele}\ \emph {et~al.}(2009)\citenamefont {Thiele},
  \citenamefont {Goyeau},\ and\ \citenamefont {Velarde}}]{ThGV2009pf}%
  \BibitemOpen
  \bibfield  {author} {\bibinfo {author} {\bibfnamefont {U.}~\bibnamefont
  {Thiele}}, \bibinfo {author} {\bibfnamefont {B.}~\bibnamefont {Goyeau}},\
  and\ \bibinfo {author} {\bibfnamefont {M.~G.}\ \bibnamefont {Velarde}},\
  }\bibfield  {title} {\bibinfo {title} {Film flow on a porous substrate},\
  }\href {https://doi.org/10.1063/1.3054157} {\bibfield  {journal} {\bibinfo
  {journal} {Phys. Fluids}\ }\textbf {\bibinfo {volume} {21}},\ \bibinfo
  {pages} {014103} (\bibinfo {year} {2009})}\BibitemShut {NoStop}%
\bibitem [{\citenamefont {Knox}\ \emph {et~al.}(2015)\citenamefont {Knox},
  \citenamefont {Wilson}, \citenamefont {Duffy},\ and\ \citenamefont
  {Mckee}}]{KWDM2015ijam}%
  \BibitemOpen
  \bibfield  {author} {\bibinfo {author} {\bibfnamefont {D.~J.}\ \bibnamefont
  {Knox}}, \bibinfo {author} {\bibfnamefont {S.~K.}\ \bibnamefont {Wilson}},
  \bibinfo {author} {\bibfnamefont {B.~R.}\ \bibnamefont {Duffy}},\ and\
  \bibinfo {author} {\bibfnamefont {S.}~\bibnamefont {Mckee}},\ }\bibfield
  {title} {\bibinfo {title} {Porous squeeze-film flow},\ }\href
  {https://doi.org/10.1093/imamat/hxt042} {\bibfield  {journal} {\bibinfo
  {journal} {IMA J. Appl. Math.}\ }\textbf {\bibinfo {volume} {80}},\ \bibinfo
  {pages} {376} (\bibinfo {year} {2015})}\BibitemShut {NoStop}%
\bibitem [{\citenamefont {Angot}\ \emph {et~al.}(2017)\citenamefont {Angot},
  \citenamefont {Goyeau},\ and\ \citenamefont {Ochoa-Tapia}}]{AnGO2017pre}%
  \BibitemOpen
  \bibfield  {author} {\bibinfo {author} {\bibfnamefont {P.}~\bibnamefont
  {Angot}}, \bibinfo {author} {\bibfnamefont {B.}~\bibnamefont {Goyeau}},\ and\
  \bibinfo {author} {\bibfnamefont {J.~A.}\ \bibnamefont {Ochoa-Tapia}},\
  }\bibfield  {title} {\bibinfo {title} {Asymptotic modeling of transport
  phenomena at the interface between a fluid and a porous layer: jump
  conditions},\ }\href {https://doi.org/10.1103/PhysRevE.95.063302} {\bibfield
  {journal} {\bibinfo  {journal} {Phys. Rev. E}\ }\textbf {\bibinfo {volume}
  {95}},\ \bibinfo {pages} {063302} (\bibinfo {year} {2017})}\BibitemShut
  {NoStop}%
\bibitem [{\citenamefont {Fitzgerald}\ and\ \citenamefont
  {Woods}(1994)}]{FiWo1994n}%
  \BibitemOpen
  \bibfield  {author} {\bibinfo {author} {\bibfnamefont {S.~D.}\ \bibnamefont
  {Fitzgerald}}\ and\ \bibinfo {author} {\bibfnamefont {A.~W.}\ \bibnamefont
  {Woods}},\ }\bibfield  {title} {\bibinfo {title} {The instability of a
  vaporization front in hot porous rock},\ }\href
  {https://doi.org/10.1038/367450a0} {\bibfield  {journal} {\bibinfo  {journal}
  {Nature}\ }\textbf {\bibinfo {volume} {367}},\ \bibinfo {pages} {450}
  (\bibinfo {year} {1994})}\BibitemShut {NoStop}%
\bibitem [{\citenamefont {Angelopoulos}\ \emph {et~al.}(1998)\citenamefont
  {Angelopoulos}, \citenamefont {Paunov}, \citenamefont {Burganos},\ and\
  \citenamefont {Payatakes}}]{APBP1998pre}%
  \BibitemOpen
  \bibfield  {author} {\bibinfo {author} {\bibfnamefont {A.~D.}\ \bibnamefont
  {Angelopoulos}}, \bibinfo {author} {\bibfnamefont {V.~N.}\ \bibnamefont
  {Paunov}}, \bibinfo {author} {\bibfnamefont {V.~N.}\ \bibnamefont
  {Burganos}},\ and\ \bibinfo {author} {\bibfnamefont {A.~C.}\ \bibnamefont
  {Payatakes}},\ }\bibfield  {title} {\bibinfo {title} {Lattice {B}oltzmann
  simulation of nonideal vapor-liquid flow in porous media},\ }\href
  {https://doi.org/10.1103/PhysRevE.57.3237} {\bibfield  {journal} {\bibinfo
  {journal} {Phys. Rev. E}\ }\textbf {\bibinfo {volume} {57}},\ \bibinfo
  {pages} {3237} (\bibinfo {year} {1998})}\BibitemShut {NoStop}%
\bibitem [{\citenamefont {Cirillo}\ \emph {et~al.}(2010)\citenamefont
  {Cirillo}, \citenamefont {Ianiro},\ and\ \citenamefont
  {Sciarra}}]{CiIS2010pre}%
  \BibitemOpen
  \bibfield  {author} {\bibinfo {author} {\bibfnamefont {E.~N.~M.}\
  \bibnamefont {Cirillo}}, \bibinfo {author} {\bibfnamefont {N.}~\bibnamefont
  {Ianiro}},\ and\ \bibinfo {author} {\bibfnamefont {G.}~\bibnamefont
  {Sciarra}},\ }\bibfield  {title} {\bibinfo {title} {Phase coexistence in
  consolidating porous media},\ }\href
  {https://doi.org/10.1103/PhysRevE.81.061121} {\bibfield  {journal} {\bibinfo
  {journal} {Phys. Rev. E}\ }\textbf {\bibinfo {volume} {81}},\ \bibinfo
  {pages} {061121} (\bibinfo {year} {2010})}\BibitemShut {NoStop}%
\bibitem [{\citenamefont {Chen}\ \emph {et~al.}(2015)\citenamefont {Chen},
  \citenamefont {Watanabe},\ and\ \citenamefont {Yoshikawa}}]{ChWY2015jpcc}%
  \BibitemOpen
  \bibfield  {author} {\bibinfo {author} {\bibfnamefont {Y.~J.}\ \bibnamefont
  {Chen}}, \bibinfo {author} {\bibfnamefont {S.}~\bibnamefont {Watanabe}},\
  and\ \bibinfo {author} {\bibfnamefont {K.}~\bibnamefont {Yoshikawa}},\
  }\bibfield  {title} {\bibinfo {title} {Roughening dynamics of radial
  imbibition in a porous medium},\ }\href
  {https://doi.org/10.1021/acs.jpcc.5b03157} {\bibfield  {journal} {\bibinfo
  {journal} {J. Phys. Chem. C}\ }\textbf {\bibinfo {volume} {119}},\ \bibinfo
  {pages} {12508} (\bibinfo {year} {2015})}\BibitemShut {NoStop}%
\bibitem [{\citenamefont {Beltrame}\ and\ \citenamefont
  {Cajot}(2022)}]{BeCa2022el}%
  \BibitemOpen
  \bibfield  {author} {\bibinfo {author} {\bibfnamefont {P.}~\bibnamefont
  {Beltrame}}\ and\ \bibinfo {author} {\bibfnamefont {F.}~\bibnamefont
  {Cajot}},\ }\bibfield  {title} {\bibinfo {title} {Model of hydrophobic porous
  media applied to stratified media: {W}ater trapping, intermittent flow and
  fingering instability},\ }\href {https://doi.org/10.1209/0295-5075/ac71c1}
  {\bibfield  {journal} {\bibinfo  {journal} {Europhys Lett}\ }\textbf
  {\bibinfo {volume} {138}},\ \bibinfo {pages} {53004} (\bibinfo {year}
  {2022})}\BibitemShut {NoStop}%
\bibitem [{\citenamefont {de~Wit}\ and\ \citenamefont
  {Homsy}(1999)}]{WiHo1999pf}%
  \BibitemOpen
  \bibfield  {author} {\bibinfo {author} {\bibfnamefont {A.}~\bibnamefont
  {de~Wit}}\ and\ \bibinfo {author} {\bibfnamefont {G.~M.}\ \bibnamefont
  {Homsy}},\ }\bibfield  {title} {\bibinfo {title} {Nonlinear interactions of
  chemical reactions and viscous fingering in porous media},\ }\href@noop {}
  {\bibfield  {journal} {\bibinfo  {journal} {Phys. Fluids}\ }\textbf {\bibinfo
  {volume} {11}},\ \bibinfo {pages} {949} (\bibinfo {year} {1999})}\BibitemShut
  {NoStop}%
\bibitem [{\citenamefont {De~Wit}(2001)}]{DeW2001prl}%
  \BibitemOpen
  \bibfield  {author} {\bibinfo {author} {\bibfnamefont {A.}~\bibnamefont
  {De~Wit}},\ }\bibfield  {title} {\bibinfo {title} {Fingering of chemical
  fronts in porous media},\ }\href@noop {} {\bibfield  {journal} {\bibinfo
  {journal} {Phys. Rev. Lett.}\ }\textbf {\bibinfo {volume} {8705}},\ \bibinfo
  {pages} {054502} (\bibinfo {year} {2001})}\BibitemShut {NoStop}%
\bibitem [{\citenamefont {Do}\ \emph {et~al.}(2003)\citenamefont {Do},
  \citenamefont {Do},\ and\ \citenamefont {Tran}}]{DoDT2003l}%
  \BibitemOpen
  \bibfield  {author} {\bibinfo {author} {\bibfnamefont {D.~D.}\ \bibnamefont
  {Do}}, \bibinfo {author} {\bibfnamefont {H.~D.}\ \bibnamefont {Do}},\ and\
  \bibinfo {author} {\bibfnamefont {K.~N.}\ \bibnamefont {Tran}},\ }\bibfield
  {title} {\bibinfo {title} {Analysis of adsorption of gases and vapors on
  nonporous graphitized thermal carbon black},\ }\href
  {https://doi.org/10.1021/la020191e} {\bibfield  {journal} {\bibinfo
  {journal} {Langmuir}\ }\textbf {\bibinfo {volume} {19}},\ \bibinfo {pages}
  {5656} (\bibinfo {year} {2003})}\BibitemShut {NoStop}%
\bibitem [{\citenamefont {Samaha}\ and\ \citenamefont {Gad-el
  Hak}(2014)}]{SaGa2014p}%
  \BibitemOpen
  \bibfield  {author} {\bibinfo {author} {\bibfnamefont {M.~A.}\ \bibnamefont
  {Samaha}}\ and\ \bibinfo {author} {\bibfnamefont {M.}~\bibnamefont {Gad-el
  Hak}},\ }\bibfield  {title} {\bibinfo {title} {Polymeric slippery coatings:
  {N}ature and applications},\ }\href {https://doi.org/10.3390/polym6051266}
  {\bibfield  {journal} {\bibinfo  {journal} {Polymers}\ }\textbf {\bibinfo
  {volume} {6}},\ \bibinfo {pages} {1266} (\bibinfo {year} {2014})}\BibitemShut
  {NoStop}%
\bibitem [{\citenamefont {Guan}\ \emph {et~al.}(2015)\citenamefont {Guan},
  \citenamefont {Wells}, \citenamefont {Xu}, \citenamefont {McHale},
  \citenamefont {Wood}, \citenamefont {Martin},\ and\ \citenamefont
  {Stuart-Cole}}]{GWXM2015l}%
  \BibitemOpen
  \bibfield  {author} {\bibinfo {author} {\bibfnamefont {J.~H.}\ \bibnamefont
  {Guan}}, \bibinfo {author} {\bibfnamefont {G.~G.}\ \bibnamefont {Wells}},
  \bibinfo {author} {\bibfnamefont {B.}~\bibnamefont {Xu}}, \bibinfo {author}
  {\bibfnamefont {G.}~\bibnamefont {McHale}}, \bibinfo {author} {\bibfnamefont
  {D.}~\bibnamefont {Wood}}, \bibinfo {author} {\bibfnamefont {J.}~\bibnamefont
  {Martin}},\ and\ \bibinfo {author} {\bibfnamefont {S.}~\bibnamefont
  {Stuart-Cole}},\ }\bibfield  {title} {\bibinfo {title} {Evaporation of
  sessile droplets on slippery liquid-infused porous surfaces (slips)},\ }\href
  {https://doi.org/10.1021/acs.langmuir.5b03240} {\bibfield  {journal}
  {\bibinfo  {journal} {Langmuir}\ }\textbf {\bibinfo {volume} {31}},\ \bibinfo
  {pages} {11781} (\bibinfo {year} {2015})}\BibitemShut {NoStop}%
\bibitem [{\citenamefont {Manabe}\ \emph {et~al.}(2016)\citenamefont {Manabe},
  \citenamefont {Matsubayashi}, \citenamefont {Tenjimbayashi}, \citenamefont
  {Moriya}, \citenamefont {Tsuge}, \citenamefont {Kyung},\ and\ \citenamefont
  {Shiratori}}]{MMTM2016an}%
  \BibitemOpen
  \bibfield  {author} {\bibinfo {author} {\bibfnamefont {K.}~\bibnamefont
  {Manabe}}, \bibinfo {author} {\bibfnamefont {T.}~\bibnamefont
  {Matsubayashi}}, \bibinfo {author} {\bibfnamefont {M.}~\bibnamefont
  {Tenjimbayashi}}, \bibinfo {author} {\bibfnamefont {T.}~\bibnamefont
  {Moriya}}, \bibinfo {author} {\bibfnamefont {Y.}~\bibnamefont {Tsuge}},
  \bibinfo {author} {\bibfnamefont {K.~H.}\ \bibnamefont {Kyung}},\ and\
  \bibinfo {author} {\bibfnamefont {S.}~\bibnamefont {Shiratori}},\ }\bibfield
  {title} {\bibinfo {title} {Controllable broadband optical transparency and
  wettability switching of temperature activated solid/liquid-infused
  nanofibrous membranes},\ }\href {https://doi.org/10.1021/acsnano.6b04333}
  {\bibfield  {journal} {\bibinfo  {journal} {ACS Nano}\ }\textbf {\bibinfo
  {volume} {10}},\ \bibinfo {pages} {9387} (\bibinfo {year}
  {2016})}\BibitemShut {NoStop}%
\bibitem [{\citenamefont {Luo}\ \emph {et~al.}(2017)\citenamefont {Luo},
  \citenamefont {Geraldi}, \citenamefont {Guan}, \citenamefont {McHale},
  \citenamefont {Wells},\ and\ \citenamefont {Fu}}]{LGGM2017pra}%
  \BibitemOpen
  \bibfield  {author} {\bibinfo {author} {\bibfnamefont {J.~T.}\ \bibnamefont
  {Luo}}, \bibinfo {author} {\bibfnamefont {N.~R.}\ \bibnamefont {Geraldi}},
  \bibinfo {author} {\bibfnamefont {J.~H.}\ \bibnamefont {Guan}}, \bibinfo
  {author} {\bibfnamefont {G.}~\bibnamefont {McHale}}, \bibinfo {author}
  {\bibfnamefont {G.~G.}\ \bibnamefont {Wells}},\ and\ \bibinfo {author}
  {\bibfnamefont {Y.~Q.}\ \bibnamefont {Fu}},\ }\bibfield  {title} {\bibinfo
  {title} {Slippery liquid-infused porous surfaces and droplet transportation
  by surface acoustic waves},\ }\href
  {https://doi.org/10.1103/PhysRevApplied.7.014017} {\bibfield  {journal}
  {\bibinfo  {journal} {Phys. Rev. Appl.}\ }\textbf {\bibinfo {volume} {7}},\
  \bibinfo {pages} {014017} (\bibinfo {year} {2017})}\BibitemShut {NoStop}%
\bibitem [{\citenamefont {Brabcova}\ \emph {et~al.}(2017)\citenamefont
  {Brabcova}, \citenamefont {McHale}, \citenamefont {Wells}, \citenamefont
  {Brown},\ and\ \citenamefont {Newton}}]{BMWB2017apl}%
  \BibitemOpen
  \bibfield  {author} {\bibinfo {author} {\bibfnamefont {Z.}~\bibnamefont
  {Brabcova}}, \bibinfo {author} {\bibfnamefont {G.}~\bibnamefont {McHale}},
  \bibinfo {author} {\bibfnamefont {G.~G.}\ \bibnamefont {Wells}}, \bibinfo
  {author} {\bibfnamefont {C.~V.}\ \bibnamefont {Brown}},\ and\ \bibinfo
  {author} {\bibfnamefont {M.~I.}\ \bibnamefont {Newton}},\ }\bibfield  {title}
  {\bibinfo {title} {Electric field induced reversible spreading of droplets
  into films on lubricant impregnated surfaces},\ }\href
  {https://doi.org/10.1063/1.4978859} {\bibfield  {journal} {\bibinfo
  {journal} {Appl. Phys. Lett.}\ }\textbf {\bibinfo {volume} {110}},\ \bibinfo
  {pages} {121603} (\bibinfo {year} {2017})}\BibitemShut {NoStop}%
\bibitem [{\citenamefont {Thiele}(2010)}]{Thie2010jpcm}%
  \BibitemOpen
  \bibfield  {author} {\bibinfo {author} {\bibfnamefont {U.}~\bibnamefont
  {Thiele}},\ }\bibfield  {title} {\bibinfo {title} {Thin film evolution
  equations from (evaporating) dewetting liquid layers to epitaxial growth},\
  }\href {https://doi.org/10.1088/0953-8984/22/8/084019} {\bibfield  {journal}
  {\bibinfo  {journal} {J. Phys.: Condens. Matter}\ }\textbf {\bibinfo {volume}
  {22}},\ \bibinfo {pages} {084019} (\bibinfo {year} {2010})}\BibitemShut
  {NoStop}%
\bibitem [{\citenamefont {Cueto-Felgueroso}\ and\ \citenamefont
  {Juanes}(2009{\natexlab{c}})}]{CuJu2009jcp}%
  \BibitemOpen
  \bibfield  {author} {\bibinfo {author} {\bibfnamefont {L.}~\bibnamefont
  {Cueto-Felgueroso}}\ and\ \bibinfo {author} {\bibfnamefont {R.}~\bibnamefont
  {Juanes}},\ }\bibfield  {title} {\bibinfo {title} {Adaptive rational spectral
  methods for the linear stability analysis of nonlinear fourth-order
  problems},\ }\href {https://doi.org/10.1016/j.jcp.2009.05.045} {\bibfield
  {journal} {\bibinfo  {journal} {J. Comput. Phys.}\ }\textbf {\bibinfo
  {volume} {228}},\ \bibinfo {pages} {6536} (\bibinfo {year}
  {2009}{\natexlab{c}})}\BibitemShut {NoStop}%
\bibitem [{\citenamefont {Cueto-Felgueroso}\ and\ \citenamefont
  {Juanes}(2010)}]{CuJu2010wrr}%
  \BibitemOpen
  \bibfield  {author} {\bibinfo {author} {\bibfnamefont {L.}~\bibnamefont
  {Cueto-Felgueroso}}\ and\ \bibinfo {author} {\bibfnamefont {R.}~\bibnamefont
  {Juanes}},\ }\bibfield  {title} {\bibinfo {title} {Reply to comment by david
  a. dicarlo on "a phase field model of unsaturated flow"},\ }\href
  {https://doi.org/10.1029/2010WR009723} {\bibfield  {journal} {\bibinfo
  {journal} {Water Resour. Res.}\ }\textbf {\bibinfo {volume} {46}},\ \bibinfo
  {pages} {W12802} (\bibinfo {year} {2010})}\BibitemShut {NoStop}%
\bibitem [{\citenamefont {Kap}\ \emph {et~al.}(2023)\citenamefont {Kap},
  \citenamefont {Hartmann}, \citenamefont {Hoek}, \citenamefont {de~Beer},
  \citenamefont {Siretanu}, \citenamefont {Thiele},\ and\ \citenamefont
  {Mugele}}]{KHHB2023jcp}%
  \BibitemOpen
  \bibfield  {author} {\bibinfo {author} {\bibfnamefont {{\"O}.}~\bibnamefont
  {Kap}}, \bibinfo {author} {\bibfnamefont {S.}~\bibnamefont {Hartmann}},
  \bibinfo {author} {\bibfnamefont {H.}~\bibnamefont {Hoek}}, \bibinfo {author}
  {\bibfnamefont {S.}~\bibnamefont {de~Beer}}, \bibinfo {author} {\bibfnamefont
  {I.}~\bibnamefont {Siretanu}}, \bibinfo {author} {\bibfnamefont
  {U.}~\bibnamefont {Thiele}},\ and\ \bibinfo {author} {\bibfnamefont
  {F.}~\bibnamefont {Mugele}},\ }\bibfield  {title} {\bibinfo {title}
  {Nonequilibrium configurations of swelling polymer brush layers induced by
  spreading drops of weakly volatile oil},\ }\href
  {https://doi.org/10.1063/5.0146779} {\bibfield  {journal} {\bibinfo
  {journal} {J. Chem. Phys.}\ }\textbf {\bibinfo {volume} {158}},\ \bibinfo
  {pages} {174903} (\bibinfo {year} {2023})}\BibitemShut {NoStop}%
\bibitem [{\citenamefont {Hartmann}\ \emph {et~al.}(2024)\citenamefont
  {Hartmann}, \citenamefont {Diekmann}, \citenamefont {Greve},\ and\
  \citenamefont {Thiele}}]{HDGT2024l}%
  \BibitemOpen
  \bibfield  {author} {\bibinfo {author} {\bibfnamefont {S.}~\bibnamefont
  {Hartmann}}, \bibinfo {author} {\bibfnamefont {J.}~\bibnamefont {Diekmann}},
  \bibinfo {author} {\bibfnamefont {D.}~\bibnamefont {Greve}},\ and\ \bibinfo
  {author} {\bibfnamefont {U.}~\bibnamefont {Thiele}},\ }\bibfield  {title}
  {\bibinfo {title} {Drops on polymer brushes -- {A}dvances in thin-film
  modelling of adaptive substrates},\ }\href
  {https://doi.org/10.1021/acs.langmuir.3c03313} {\bibfield  {journal}
  {\bibinfo  {journal} {Langmuir}\ }\textbf {\bibinfo {volume} {40}},\ \bibinfo
  {pages} {4001} (\bibinfo {year} {2024})}\BibitemShut {NoStop}%
\bibitem [{\citenamefont {Cahn}(1961)}]{Cahn1961am}%
  \BibitemOpen
  \bibfield  {author} {\bibinfo {author} {\bibfnamefont {J.~W.}\ \bibnamefont
  {Cahn}},\ }\bibfield  {title} {\bibinfo {title} {On spinodal decomposition},\
  }\href {https://doi.org/10.1016/0001-6160(61)90182-1} {\bibfield  {journal}
  {\bibinfo  {journal} {Acta Met.}\ }\textbf {\bibinfo {volume} {9}},\ \bibinfo
  {pages} {795} (\bibinfo {year} {1961})}\BibitemShut {NoStop}%
\bibitem [{\citenamefont {Cahn}(1965)}]{Cahn1965jcp}%
  \BibitemOpen
  \bibfield  {author} {\bibinfo {author} {\bibfnamefont {J.~W.}\ \bibnamefont
  {Cahn}},\ }\bibfield  {title} {\bibinfo {title} {Phase separation by spinodal
  decomposition in isotropic systems},\ }\href
  {https://doi.org/10.1063/1.1695731} {\bibfield  {journal} {\bibinfo
  {journal} {J. Chem. Phys.}\ }\textbf {\bibinfo {volume} {42}},\ \bibinfo
  {pages} {93} (\bibinfo {year} {1965})}\BibitemShut {NoStop}%
\bibitem [{\citenamefont {Bray}(1994)}]{Bray1994ap}%
  \BibitemOpen
  \bibfield  {author} {\bibinfo {author} {\bibfnamefont {A.~J.}\ \bibnamefont
  {Bray}},\ }\bibfield  {title} {\bibinfo {title} {Theory of phase-ordering
  kinetics},\ }\href {https://doi.org/10.1080/00018739400101505} {\bibfield
  {journal} {\bibinfo  {journal} {Adv. Phys.}\ }\textbf {\bibinfo {volume}
  {43}},\ \bibinfo {pages} {357} (\bibinfo {year} {1994})}\BibitemShut
  {NoStop}%
\bibitem [{\citenamefont {Mitlin}(1993)}]{Mitl1993jcis}%
  \BibitemOpen
  \bibfield  {author} {\bibinfo {author} {\bibfnamefont {V.~S.}\ \bibnamefont
  {Mitlin}},\ }\bibfield  {title} {\bibinfo {title} {Dewetting of solid
  surface: {A}nalogy with spinodal decomposition},\ }\href
  {https://doi.org/10.1006/jcis.1993.1142} {\bibfield  {journal} {\bibinfo
  {journal} {J. Colloid Interface Sci.}\ }\textbf {\bibinfo {volume} {156}},\
  \bibinfo {pages} {491} (\bibinfo {year} {1993})}\BibitemShut {NoStop}%
\bibitem [{\citenamefont {Oron}\ \emph {et~al.}(1997)\citenamefont {Oron},
  \citenamefont {Davis},\ and\ \citenamefont {Bankoff}}]{OrDB1997rmp}%
  \BibitemOpen
  \bibfield  {author} {\bibinfo {author} {\bibfnamefont {A.}~\bibnamefont
  {Oron}}, \bibinfo {author} {\bibfnamefont {S.~H.}\ \bibnamefont {Davis}},\
  and\ \bibinfo {author} {\bibfnamefont {S.~G.}\ \bibnamefont {Bankoff}},\
  }\bibfield  {title} {\bibinfo {title} {Long-scale evolution of thin liquid
  films},\ }\href {https://doi.org/10.1103/RevModPhys.69.931} {\bibfield
  {journal} {\bibinfo  {journal} {Rev. Mod. Phys.}\ }\textbf {\bibinfo {volume}
  {69}},\ \bibinfo {pages} {931} (\bibinfo {year} {1997})}\BibitemShut
  {NoStop}%
\bibitem [{\citenamefont {Pototsky}\ \emph {et~al.}(2005)\citenamefont
  {Pototsky}, \citenamefont {Bestehorn}, \citenamefont {Merkt},\ and\
  \citenamefont {Thiele}}]{PBMT2005jcp}%
  \BibitemOpen
  \bibfield  {author} {\bibinfo {author} {\bibfnamefont {A.}~\bibnamefont
  {Pototsky}}, \bibinfo {author} {\bibfnamefont {M.}~\bibnamefont {Bestehorn}},
  \bibinfo {author} {\bibfnamefont {D.}~\bibnamefont {Merkt}},\ and\ \bibinfo
  {author} {\bibfnamefont {U.}~\bibnamefont {Thiele}},\ }\bibfield  {title}
  {\bibinfo {title} {Morphology changes in the evolution of liquid two-layer
  films},\ }\href {https://doi.org/10.1063/1.1927512} {\bibfield  {journal}
  {\bibinfo  {journal} {J. Chem. Phys.}\ }\textbf {\bibinfo {volume} {122}},\
  \bibinfo {pages} {224711} (\bibinfo {year} {2005})}\BibitemShut {NoStop}%
\bibitem [{\citenamefont {Bommer}\ \emph {et~al.}(2013)\citenamefont {Bommer},
  \citenamefont {Cartellier}, \citenamefont {Jachalski}, \citenamefont
  {Peschka}, \citenamefont {Seemann},\ and\ \citenamefont
  {Wagner}}]{BCJP2013epje}%
  \BibitemOpen
  \bibfield  {author} {\bibinfo {author} {\bibfnamefont {S.}~\bibnamefont
  {Bommer}}, \bibinfo {author} {\bibfnamefont {F.}~\bibnamefont {Cartellier}},
  \bibinfo {author} {\bibfnamefont {S.}~\bibnamefont {Jachalski}}, \bibinfo
  {author} {\bibfnamefont {D.}~\bibnamefont {Peschka}}, \bibinfo {author}
  {\bibfnamefont {R.}~\bibnamefont {Seemann}},\ and\ \bibinfo {author}
  {\bibfnamefont {B.}~\bibnamefont {Wagner}},\ }\bibfield  {title} {\bibinfo
  {title} {Droplets on liquids and their journey into equilibrium},\ }\href
  {https://doi.org/10.1140/epje/i2013-13087-x} {\bibfield  {journal} {\bibinfo
  {journal} {Eur. Phys. J. E}\ }\textbf {\bibinfo {volume} {36}},\ \bibinfo
  {pages} {87} (\bibinfo {year} {2013})}\BibitemShut {NoStop}%
\bibitem [{\citenamefont {Thiele}\ \emph {et~al.}(2013)\citenamefont {Thiele},
  \citenamefont {Todorova},\ and\ \citenamefont {Lopez}}]{ThTL2013prl}%
  \BibitemOpen
  \bibfield  {author} {\bibinfo {author} {\bibfnamefont {U.}~\bibnamefont
  {Thiele}}, \bibinfo {author} {\bibfnamefont {D.~V.}\ \bibnamefont
  {Todorova}},\ and\ \bibinfo {author} {\bibfnamefont {H.}~\bibnamefont
  {Lopez}},\ }\bibfield  {title} {\bibinfo {title} {Gradient dynamics
  description for films of mixtures and suspensions: dewetting triggered by
  coupled film height and concentration fluctuations},\ }\href
  {https://doi.org/10.1103/PhysRevLett.111.117801} {\bibfield  {journal}
  {\bibinfo  {journal} {Phys. Rev. Lett.}\ }\textbf {\bibinfo {volume} {111}},\
  \bibinfo {pages} {117801} (\bibinfo {year} {2013})}\BibitemShut {NoStop}%
\bibitem [{\citenamefont {Thiele}\ \emph {et~al.}(2016)\citenamefont {Thiele},
  \citenamefont {Archer},\ and\ \citenamefont {Pismen}}]{ThAP2016prf}%
  \BibitemOpen
  \bibfield  {author} {\bibinfo {author} {\bibfnamefont {U.}~\bibnamefont
  {Thiele}}, \bibinfo {author} {\bibfnamefont {A.~J.}\ \bibnamefont {Archer}},\
  and\ \bibinfo {author} {\bibfnamefont {L.~M.}\ \bibnamefont {Pismen}},\
  }\bibfield  {title} {\bibinfo {title} {Gradient dynamics models for liquid
  films with soluble surfactant},\ }\href
  {https://doi.org/10.1103/PhysRevFluids.1.083903} {\bibfield  {journal}
  {\bibinfo  {journal} {Phys. Rev. Fluids}\ }\textbf {\bibinfo {volume} {1}},\
  \bibinfo {pages} {083903} (\bibinfo {year} {2016})}\BibitemShut {NoStop}%
\bibitem [{\citenamefont {Henkel}\ \emph {et~al.}(2022)\citenamefont {Henkel},
  \citenamefont {Essink}, \citenamefont {Hoang}, \citenamefont {van Zwieten},
  \citenamefont {van Brummelen}, \citenamefont {Thiele},\ and\ \citenamefont
  {Snoeijer}}]{HEHZ2022prsa}%
  \BibitemOpen
  \bibfield  {author} {\bibinfo {author} {\bibfnamefont {C.}~\bibnamefont
  {Henkel}}, \bibinfo {author} {\bibfnamefont {M.}~\bibnamefont {Essink}},
  \bibinfo {author} {\bibfnamefont {T.}~\bibnamefont {Hoang}}, \bibinfo
  {author} {\bibfnamefont {G.}~\bibnamefont {van Zwieten}}, \bibinfo {author}
  {\bibfnamefont {E.}~\bibnamefont {van Brummelen}}, \bibinfo {author}
  {\bibfnamefont {U.}~\bibnamefont {Thiele}},\ and\ \bibinfo {author}
  {\bibfnamefont {J.}~\bibnamefont {Snoeijer}},\ }\bibfield  {title} {\bibinfo
  {title} {Soft wetting with (a)symmetric {S}huttleworth effect},\ }\href
  {https://doi.org/10.1098/rspa.2022.0132} {\bibfield  {journal} {\bibinfo
  {journal} {Proc. R. Soc. A}\ }\textbf {\bibinfo {volume} {478}},\ \bibinfo
  {pages} {20220132} (\bibinfo {year} {2022})}\BibitemShut {NoStop}%
\bibitem [{\citenamefont {Thiele}(2018)}]{Thie2018csa}%
  \BibitemOpen
  \bibfield  {author} {\bibinfo {author} {\bibfnamefont {U.}~\bibnamefont
  {Thiele}},\ }\bibfield  {title} {\bibinfo {title} {Recent advances in and
  future challenges for mesoscopic hydrodynamic modelling of complex wetting},\
  }\href {https://doi.org/10.1016/j.colsurfa.2018.05.049} {\bibfield  {journal}
  {\bibinfo  {journal} {Colloid Surf. A}\ }\textbf {\bibinfo {volume} {553}},\
  \bibinfo {pages} {487} (\bibinfo {year} {2018})}\BibitemShut {NoStop}%
\bibitem [{\citenamefont {Hartmann}\ \emph {et~al.}(2023)\citenamefont
  {Hartmann}, \citenamefont {Diddens}, \citenamefont {Jalaal},\ and\
  \citenamefont {Thiele}}]{HDJT2023jfm}%
  \BibitemOpen
  \bibfield  {author} {\bibinfo {author} {\bibfnamefont {S.}~\bibnamefont
  {Hartmann}}, \bibinfo {author} {\bibfnamefont {C.}~\bibnamefont {Diddens}},
  \bibinfo {author} {\bibfnamefont {M.}~\bibnamefont {Jalaal}},\ and\ \bibinfo
  {author} {\bibfnamefont {U.}~\bibnamefont {Thiele}},\ }\bibfield  {title}
  {\bibinfo {title} {Sessile drop evaporation in a gap - crossover between
  diffusion-limited and phase transition-limited regime},\ }\href
  {https://doi.org/10.1017/jfm.2023.176} {\bibfield  {journal} {\bibinfo
  {journal} {J. Fluid Mech.}\ }\textbf {\bibinfo {volume} {960}},\ \bibinfo
  {pages} {A32} (\bibinfo {year} {2023})}\BibitemShut {NoStop}%
\bibitem [{\citenamefont {Thiele}(2007)}]{Thie2007chapter}%
  \BibitemOpen
  \bibfield  {author} {\bibinfo {author} {\bibfnamefont {U.}~\bibnamefont
  {Thiele}},\ }\bibfield  {title} {\bibinfo {title} {Structure formation in
  thin liquid films},\ }in\ \href
  {https://doi.org/10.1007/978-3-211-69808-2\_2} {\emph {\bibinfo {booktitle}
  {Thin Films of Soft Matter}}},\ \bibinfo {editor} {edited by\ \bibinfo
  {editor} {\bibfnamefont {S.}~\bibnamefont {Kalliadasis}}\ and\ \bibinfo
  {editor} {\bibfnamefont {U.}~\bibnamefont {Thiele}}}\ (\bibinfo  {publisher}
  {Springer Vienna},\ \bibinfo {address} {Vienna},\ \bibinfo {year} {2007})\
  pp.\ \bibinfo {pages} {25--93}\BibitemShut {NoStop}%
\bibitem [{\citenamefont {van Genuchten}(1980)}]{Genu1980sssaj}%
  \BibitemOpen
  \bibfield  {author} {\bibinfo {author} {\bibfnamefont {M.~T.}\ \bibnamefont
  {van Genuchten}},\ }\bibfield  {title} {\bibinfo {title} {A closed-form
  equation for predicting the hydraulic conductivity of unsaturated soils},\
  }\href {https://doi.org/10.2136/sssaj1980.03615995004400050002x} {\bibfield
  {journal} {\bibinfo  {journal} {Soil Sci. Soc. Am. J.}\ }\textbf {\bibinfo
  {volume} {44}},\ \bibinfo {pages} {892} (\bibinfo {year} {1980})}\BibitemShut
  {NoStop}%
\bibitem [{\citenamefont {Jurin}(1718)}]{Juri1718pt}%
  \BibitemOpen
  \bibfield  {author} {\bibinfo {author} {\bibfnamefont {J.}~\bibnamefont
  {Jurin}},\ }\bibfield  {title} {\bibinfo {title} {An account of some
  experiments shown before the {R}oyal {S}ociety; {W}ith an enquiry into the
  cause of the ascent and suspension of water in capillary tubes},\ }\href@noop
  {} {\bibfield  {journal} {\bibinfo  {journal} {Phil. Trans.}\ }\textbf
  {\bibinfo {volume} {30}},\ \bibinfo {pages} {739} (\bibinfo {year}
  {1718})}\BibitemShut {NoStop}%
\bibitem [{\citenamefont {Doi}(2013)}]{Doi2013}%
  \BibitemOpen
  \bibfield  {author} {\bibinfo {author} {\bibfnamefont {M.}~\bibnamefont
  {Doi}},\ }\href {https://doi.org/10.1007/b97416} {\emph {\bibinfo {title}
  {Soft matter physics}}}\ (\bibinfo  {publisher} {Oxford University Press},\
  \bibinfo {address} {Oxford},\ \bibinfo {year} {2013})\BibitemShut {NoStop}%
\bibitem [{\citenamefont {Greve}\ \emph {et~al.}(2023)\citenamefont {Greve},
  \citenamefont {Hartmann},\ and\ \citenamefont {Thiele}}]{GrHT2023sm}%
  \BibitemOpen
  \bibfield  {author} {\bibinfo {author} {\bibfnamefont {D.}~\bibnamefont
  {Greve}}, \bibinfo {author} {\bibfnamefont {S.}~\bibnamefont {Hartmann}},\
  and\ \bibinfo {author} {\bibfnamefont {U.}~\bibnamefont {Thiele}},\
  }\bibfield  {title} {\bibinfo {title} {Stick-slip dynamics in the forced
  wetting of polymer brushes},\ }\href {https://doi.org/10.1039/D3SM00104K}
  {\bibfield  {journal} {\bibinfo  {journal} {Soft Matter}\ }\textbf {\bibinfo
  {volume} {19}},\ \bibinfo {pages} {4041} (\bibinfo {year}
  {2023})}\BibitemShut {NoStop}%
\bibitem [{\citenamefont {Heil}\ and\ \citenamefont {Hazel}(2006)}]{HeHa2006}%
  \BibitemOpen
  \bibfield  {author} {\bibinfo {author} {\bibfnamefont {M.}~\bibnamefont
  {Heil}}\ and\ \bibinfo {author} {\bibfnamefont {A.~L.}\ \bibnamefont
  {Hazel}},\ }\bibfield  {title} {\bibinfo {title} {Oomph-lib - an
  object-oriented multi-physics finite-element library},\ }in\ \href
  {https://doi.org/10.1007/3-540-34596-5_2} {\emph {\bibinfo {booktitle}
  {Fluid-Structure Interaction: Modelling, Simulation, Optimisation}}},\
  \bibinfo {editor} {edited by\ \bibinfo {editor} {\bibfnamefont {H.-J.}\
  \bibnamefont {Bungartz}}\ and\ \bibinfo {editor} {\bibfnamefont
  {M.}~\bibnamefont {Sch{\"a}fer}}}\ (\bibinfo  {publisher} {Springer},\
  \bibinfo {address} {Berlin, Heidelberg},\ \bibinfo {year} {2006})\ pp.\
  \bibinfo {pages} {19--49}\BibitemShut {NoStop}%
\bibitem [{\citenamefont {Hartmann}(2023)}]{Hartmann2023Munster}%
  \BibitemOpen
  \bibfield  {author} {\bibinfo {author} {\bibfnamefont {S.}~\bibnamefont
  {Hartmann}},\ }\emph {\bibinfo {title} {Contact Line Dynamics on Rigid and
  Adaptive Substrates}},\ \href@noop {} {Ph.D. thesis},\ \bibinfo  {school}
  {University of M{\"u}nster}, \bibinfo {address} {M{\"u}nster} (\bibinfo
  {year} {2023})\BibitemShut {NoStop}%
\bibitem [{\citenamefont {Cajot}(2024)}]{Cajot2024Avignon}%
  \BibitemOpen
  \bibfield  {author} {\bibinfo {author} {\bibfnamefont {F.}~\bibnamefont
  {Cajot}},\ }\emph {\bibinfo {title} {Modeling water transfer in soil in the
  presence of amphiphilic matter: application to the rhizosphere}},\ \href@noop
  {} {Ph.D. thesis},\ \bibinfo  {school} {University of Avignon}, \bibinfo
  {address} {Avignon} (\bibinfo {year} {2024})\BibitemShut {NoStop}%
\bibitem [{\citenamefont {Cajot}\ \emph {et~al.}(2024)\citenamefont {Cajot},
  \citenamefont {Doussan},\ and\ \citenamefont {Beltrame}}]{CaDB2024preprint}%
  \BibitemOpen
  \bibfield  {author} {\bibinfo {author} {\bibfnamefont {F.}~\bibnamefont
  {Cajot}}, \bibinfo {author} {\bibfnamefont {C.}~\bibnamefont {Doussan}},\
  and\ \bibinfo {author} {\bibfnamefont {P.}~\bibnamefont {Beltrame}},\
  }\bibfield  {title} {\bibinfo {title} {Model of water transfer through an
  amphiphilic porous medium},\ }\href@noop {} {\bibfield  {journal} {\bibinfo
  {journal} {Preprint}\ } (\bibinfo {year} {2024})}\BibitemShut {NoStop}%
\bibitem [{\citenamefont {Clarke}\ \emph {et~al.}(2002)\citenamefont {Clarke},
  \citenamefont {Blake}, \citenamefont {Carruthers},\ and\ \citenamefont
  {Woodward}}]{CBCW2002l}%
  \BibitemOpen
  \bibfield  {author} {\bibinfo {author} {\bibfnamefont {A.}~\bibnamefont
  {Clarke}}, \bibinfo {author} {\bibfnamefont {T.~D.}\ \bibnamefont {Blake}},
  \bibinfo {author} {\bibfnamefont {K.}~\bibnamefont {Carruthers}},\ and\
  \bibinfo {author} {\bibfnamefont {A.}~\bibnamefont {Woodward}},\ }\bibfield
  {title} {\bibinfo {title} {Spreading and imbibition of liquid droplets on
  porous surfaces},\ }\href {https://doi.org/10.1021/la0117810} {\bibfield
  {journal} {\bibinfo  {journal} {Langmuir}\ }\textbf {\bibinfo {volume}
  {18}},\ \bibinfo {pages} {2980} (\bibinfo {year} {2002})}\BibitemShut
  {NoStop}%
\bibitem [{\citenamefont {Bhattacharjee}\ \emph {et~al.}(2021)\citenamefont
  {Bhattacharjee}, \citenamefont {Nazaripoor}, \citenamefont {Soltannia},
  \citenamefont {Ismail},\ and\ \citenamefont {Sadrzadeh}}]{BNSI2021csaea}%
  \BibitemOpen
  \bibfield  {author} {\bibinfo {author} {\bibfnamefont {D.}~\bibnamefont
  {Bhattacharjee}}, \bibinfo {author} {\bibfnamefont {H.}~\bibnamefont
  {Nazaripoor}}, \bibinfo {author} {\bibfnamefont {B.}~\bibnamefont
  {Soltannia}}, \bibinfo {author} {\bibfnamefont {M.~F.}\ \bibnamefont
  {Ismail}},\ and\ \bibinfo {author} {\bibfnamefont {M.}~\bibnamefont
  {Sadrzadeh}},\ }\bibfield  {title} {\bibinfo {title} {An experimental and
  numerical study of droplet spreading and imbibition on microporous
  membranes},\ }\href {https://doi.org/10.1016/j.colsurfa.2021.126191}
  {\bibfield  {journal} {\bibinfo  {journal} {Colloid Surf. A-Physicochem. Eng.
  Asp.}\ }\textbf {\bibinfo {volume} {615}},\ \bibinfo {pages} {126191}
  (\bibinfo {year} {2021})}\BibitemShut {NoStop}%
\bibitem [{\citenamefont {Gimenez}\ \emph {et~al.}(2020)\citenamefont
  {Gimenez}, \citenamefont {Mercuri}, \citenamefont {Berli},\ and\
  \citenamefont {Bellino}}]{GMBB2020pccp}%
  \BibitemOpen
  \bibfield  {author} {\bibinfo {author} {\bibfnamefont {R.}~\bibnamefont
  {Gimenez}}, \bibinfo {author} {\bibfnamefont {M.}~\bibnamefont {Mercuri}},
  \bibinfo {author} {\bibfnamefont {C.~L.~A.}\ \bibnamefont {Berli}},\ and\
  \bibinfo {author} {\bibfnamefont {M.~G.}\ \bibnamefont {Bellino}},\
  }\bibfield  {title} {\bibinfo {title} {Sliding of drops on mesoporous thin
  films},\ }\href {https://doi.org/10.1039/c9cp06993c} {\bibfield  {journal}
  {\bibinfo  {journal} {Phys. Chem. Chem. Phys.}\ }\textbf {\bibinfo {volume}
  {22}},\ \bibinfo {pages} {5915} (\bibinfo {year} {2020})}\BibitemShut
  {NoStop}%
\bibitem [{\citenamefont {Starov}(2004)}]{Star2004acis}%
  \BibitemOpen
  \bibfield  {author} {\bibinfo {author} {\bibfnamefont {V.~M.}\ \bibnamefont
  {Starov}},\ }\bibfield  {title} {\bibinfo {title} {Surfactant solutions and
  porous substrates: spreading and imbibition},\ }\href
  {https://doi.org/10.1016/j.cis.2004.07.007} {\bibfield  {journal} {\bibinfo
  {journal} {Adv. Colloid Interface Sci.}\ }\textbf {\bibinfo {volume} {111}},\
  \bibinfo {pages} {3} (\bibinfo {year} {2004})}\BibitemShut {NoStop}%
\bibitem [{\citenamefont {Holman}\ \emph {et~al.}(2002)\citenamefont {Holman},
  \citenamefont {Cima}, \citenamefont {Uhland},\ and\ \citenamefont
  {Sachs}}]{HCUS2002jcis}%
  \BibitemOpen
  \bibfield  {author} {\bibinfo {author} {\bibfnamefont {R.~K.}\ \bibnamefont
  {Holman}}, \bibinfo {author} {\bibfnamefont {M.~J.}\ \bibnamefont {Cima}},
  \bibinfo {author} {\bibfnamefont {S.~A.}\ \bibnamefont {Uhland}},\ and\
  \bibinfo {author} {\bibfnamefont {E.}~\bibnamefont {Sachs}},\ }\bibfield
  {title} {\bibinfo {title} {Spreading and infiltration of inkjet-printed
  polymer solution droplets on a porous substrate},\ }\href
  {https://doi.org/10.1006/jcis.2002.8225} {\bibfield  {journal} {\bibinfo
  {journal} {J. Colloid Interface Sci.}\ }\textbf {\bibinfo {volume} {249}},\
  \bibinfo {pages} {432} (\bibinfo {year} {2002})}\BibitemShut {NoStop}%
\bibitem [{\citenamefont {von Borries~Lopes}\ \emph {et~al.}(2018)\citenamefont
  {von Borries~Lopes}, \citenamefont {Thiele},\ and\ \citenamefont
  {Hazel}}]{BoTH2018jfm}%
  \BibitemOpen
  \bibfield  {author} {\bibinfo {author} {\bibfnamefont {A.}~\bibnamefont {von
  Borries~Lopes}}, \bibinfo {author} {\bibfnamefont {U.}~\bibnamefont
  {Thiele}},\ and\ \bibinfo {author} {\bibfnamefont {A.~L.}\ \bibnamefont
  {Hazel}},\ }\bibfield  {title} {\bibinfo {title} {On the multiple solutions
  of coating and rimming flows on rotating cylinders},\ }\href
  {https://doi.org/10.1017/jfm.2017.756} {\bibfield  {journal} {\bibinfo
  {journal} {J. Fluid Mech.}\ }\textbf {\bibinfo {volume} {835}},\ \bibinfo
  {pages} {540} (\bibinfo {year} {2018})}\BibitemShut {NoStop}%
\bibitem [{\citenamefont {Diekmann}(2022)}]{Diekmann2022Munster}%
  \BibitemOpen
  \bibfield  {author} {\bibinfo {author} {\bibfnamefont {J.}~\bibnamefont
  {Diekmann}},\ }\emph {\bibinfo {title} {Macroscopic and Mesoscopic Modelling
  of Compound Drops}},\ \href@noop {} {Master's thesis},\ \bibinfo  {school}
  {University of M{\"u}nster}, \bibinfo {address} {M{\"u}nster} (\bibinfo
  {year} {2022})\BibitemShut {NoStop}%
\bibitem [{\citenamefont {Diekmann}\ and\ \citenamefont
  {Thiele}(2024{\natexlab{a}})}]{DiTh2024preprint}%
  \BibitemOpen
  \bibfield  {author} {\bibinfo {author} {\bibfnamefont {J.}~\bibnamefont
  {Diekmann}}\ and\ \bibinfo {author} {\bibfnamefont {U.}~\bibnamefont
  {Thiele}},\ }\bibfield  {title} {\bibinfo {title} {Mesoscopic hydrodynamic
  model for spreading, sliding and coarsening compound drops},\ }\href@noop {}
  {\bibfield  {journal} {\bibinfo  {journal} {Preprint}\ } (\bibinfo {year}
  {2024}{\natexlab{a}})}\BibitemShut {NoStop}%
\bibitem [{\citenamefont {Thiele}\ \emph {et~al.}(2003)\citenamefont {Thiele},
  \citenamefont {Brusch}, \citenamefont {Bestehorn},\ and\ \citenamefont
  {B{\"a}r}}]{TBBB2003epje}%
  \BibitemOpen
  \bibfield  {author} {\bibinfo {author} {\bibfnamefont {U.}~\bibnamefont
  {Thiele}}, \bibinfo {author} {\bibfnamefont {L.}~\bibnamefont {Brusch}},
  \bibinfo {author} {\bibfnamefont {M.}~\bibnamefont {Bestehorn}},\ and\
  \bibinfo {author} {\bibfnamefont {M.}~\bibnamefont {B{\"a}r}},\ }\bibfield
  {title} {\bibinfo {title} {Modelling thin-film dewetting on structured
  substrates and templates: {B}ifurcation analysis and numerical simulations},\
  }\href {https://doi.org/10.1140/epje/i2003-10019-5} {\bibfield  {journal}
  {\bibinfo  {journal} {Eur. Phys. J. E}\ }\textbf {\bibinfo {volume} {11}},\
  \bibinfo {pages} {255} (\bibinfo {year} {2003})}\BibitemShut {NoStop}%
\bibitem [{\citenamefont {Savva}\ and\ \citenamefont
  {Kalliadasis}(2012)}]{SaKa2012jem}%
  \BibitemOpen
  \bibfield  {author} {\bibinfo {author} {\bibfnamefont {N.}~\bibnamefont
  {Savva}}\ and\ \bibinfo {author} {\bibfnamefont {S.}~\bibnamefont
  {Kalliadasis}},\ }\bibfield  {title} {\bibinfo {title} {Influence of gravity
  on the spreading of two-dimensional droplets over topographical substrates},\
  }\href {https://doi.org/10.1007/s10665-010-9426-4} {\bibfield  {journal}
  {\bibinfo  {journal} {J. Eng. Math.}\ }\textbf {\bibinfo {volume} {73}},\
  \bibinfo {pages} {3} (\bibinfo {year} {2012})}\BibitemShut {NoStop}%
\bibitem [{\citenamefont {Diekmann}\ and\ \citenamefont
  {Thiele}(2024{\natexlab{b}})}]{DiTh2024epjst}%
  \BibitemOpen
  \bibfield  {author} {\bibinfo {author} {\bibfnamefont {J.}~\bibnamefont
  {Diekmann}}\ and\ \bibinfo {author} {\bibfnamefont {U.}~\bibnamefont
  {Thiele}},\ }\bibfield  {title} {\bibinfo {title} {Drops of volatile binary
  mixtures on brush-covered substrates},\ }\bibfield  {journal} {\bibinfo
  {journal} {Eur. Phys. J.-Spec. Top.}\ }\href
  {https://doi.org/10.1140/epjs/s11734-024-01169-4}
  {10.1140/epjs/s11734-024-01169-4} (\bibinfo {year}
  {2024}{\natexlab{b}})\BibitemShut {NoStop}%
\bibitem [{\citenamefont {Kvick}\ \emph {et~al.}(2017)\citenamefont {Kvick},
  \citenamefont {Martinez}, \citenamefont {Hewitt},\ and\ \citenamefont
  {Balmforth}}]{KMHB2017prf}%
  \BibitemOpen
  \bibfield  {author} {\bibinfo {author} {\bibfnamefont {M.}~\bibnamefont
  {Kvick}}, \bibinfo {author} {\bibfnamefont {D.~M.}\ \bibnamefont {Martinez}},
  \bibinfo {author} {\bibfnamefont {D.~R.}\ \bibnamefont {Hewitt}},\ and\
  \bibinfo {author} {\bibfnamefont {N.~J.}\ \bibnamefont {Balmforth}},\
  }\bibfield  {title} {\bibinfo {title} {Imbibition with swelling: capillary
  rise in thin deformable porous media},\ }\href
  {https://doi.org/10.1103/PhysRevFluids.2.074001} {\bibfield  {journal}
  {\bibinfo  {journal} {Phys. Rev. Fluids}\ }\textbf {\bibinfo {volume} {2}},\
  \bibinfo {pages} {074001} (\bibinfo {year} {2017})}\BibitemShut {NoStop}%
\bibitem [{\citenamefont {Hapgood}\ \emph {et~al.}(2002)\citenamefont
  {Hapgood}, \citenamefont {Litster}, \citenamefont {Biggs},\ and\
  \citenamefont {Howes}}]{HLBH2002jcis}%
  \BibitemOpen
  \bibfield  {author} {\bibinfo {author} {\bibfnamefont {K.~P.}\ \bibnamefont
  {Hapgood}}, \bibinfo {author} {\bibfnamefont {J.~D.}\ \bibnamefont
  {Litster}}, \bibinfo {author} {\bibfnamefont {S.~R.}\ \bibnamefont {Biggs}},\
  and\ \bibinfo {author} {\bibfnamefont {T.}~\bibnamefont {Howes}},\ }\bibfield
   {title} {\bibinfo {title} {Drop penetration into porous powder beds},\
  }\href {https://doi.org/10.1006/jcis.2002.8527} {\bibfield  {journal}
  {\bibinfo  {journal} {J. Colloid Interface Sci.}\ }\textbf {\bibinfo {volume}
  {253}},\ \bibinfo {pages} {353} (\bibinfo {year} {2002})}\BibitemShut
  {NoStop}%
\bibitem [{\citenamefont {Wooh}\ and\ \citenamefont
  {Butt}(2017)}]{WoBu2017acie}%
  \BibitemOpen
  \bibfield  {author} {\bibinfo {author} {\bibfnamefont {S.}~\bibnamefont
  {Wooh}}\ and\ \bibinfo {author} {\bibfnamefont {H.~J.}\ \bibnamefont
  {Butt}},\ }\bibfield  {title} {\bibinfo {title} {A photocatalytically active
  lubricant-impregnated surface},\ }\href
  {https://doi.org/10.1002/anie.201611277} {\bibfield  {journal} {\bibinfo
  {journal} {Angew. Chem. Int. Ed.}\ }\textbf {\bibinfo {volume} {56}},\
  \bibinfo {pages} {4965} (\bibinfo {year} {2017})}\BibitemShut {NoStop}%
\end{thebibliography}%

\end{document}